\PassOptionsToPackage{pdfpagelabels=false}{hyperref}
\documentclass[fleqn,usenatbib]{mnras}

\usepackage{newtxtext,newtxmath}
\usepackage[T1]{fontenc}

\usepackage{graphicx}	% Including figure files
\usepackage{amsmath}	% Advanced maths commands

\usepackage{amsmath}
\usepackage{setspace}

\usepackage{lineno}

\newcommand\lamyprime{\lambda^{\mathrm{new}}_{\mathrm{Y}}}
\newcommand\lamy{\lambda_{\mathrm{Y}}}
\newcommand\nsat{N_{\mathrm{mem}}}
\newcommand\lamrm{\lambda_{\mathrm{RM}}}
\newcommand\mhy{M_{180}^{L_\mathrm{tot}}}

\newcommand\mhwl{M_{h}^{\mathrm{WL}}}

\newcommand\hmsol{h^{-1}M_{\odot}}
\newcommand{\hmpc}{h^{-1}\mathrm{Mpc}}
\newcommand\redmapper{\texttt{redMaPPer}}
\newcommand\yang{\texttt{Yang}}
\newcommand\pmem{p_{\mathrm{mem}}}
\newcommand\pcen{p_{\mathrm{cen}}}

\title[Conditional Luminosity Function and Weak Lensing of Clusters]
{Satellite Content and Halo Mass of Galaxy
Clusters: Comparison between Red-Sequence and Halo-based Optical
Cluster Finders}

\author[J. B. Golden-Marx et al.]{
Jesse B. Golden-Marx,$^{1}$\thanks{E-mail: jessegm@sjtu.edu.cn }
Ying Zu,$^{1, 2, 3}$\thanks{E-mail: yingzu@sjtu.edu.cn }
Jiaqi Wang$^{1}$,
Hekun Li$^{1}$,
Jun Zhang$^{1, 2, 3}$,
Xiaohu Yang$^{1, 2, 3, 4}$
\\
$^{1}$Department of Astronomy, School of Physics and Astronomy, Shanghai Jiao Tong
University, Shanghai 200240, China\\
$^{2}$Shanghai Key Laboratory for Particle Physics and Cosmology, Shanghai Jiao Tong
University, Shanghai 200240, China\\
$^{3}$Key Laboratory for Particle Physics, Astrophysics and Cosmology,
Ministry of Education, Shanghai Jiao Tong University, Shanghai 200240,
China\\
$^{4}$Tsung-Dao Lee Institute, Shanghai Jiao Tong University, Shanghai 200240, China
}

\date{May 2023}

\pubyear{2023}

\begin{document}
\label{firstpage}
\pagerange{\pageref{firstpage}--\pageref{lastpage}}
\maketitle

\begin{abstract}
    Cluster cosmology depends critically on how optical clusters are
    selected from imaging surveys. We compare the conditional luminosity
    function (CLF) and weak lensing halo masses between two different
    cluster samples at fixed richness, detected within the same volume
    ($0.1{<}z{<}0.34$) using the red-sequence and halo-based methods.
    After calibrating our CLF deprojection method against mock galaxy
    samples, we measure the 3D CLFs by cross-correlating clusters with SDSS
    photometric galaxies. As expected, the CLFs of red-sequence and
    halo-based finders exhibit redder and bluer populations, respectively.
    The red-sequence clusters have a flat distribution of red galaxies at
    the faint end, while the halo-based clusters host a decreasing faint
    red and a boosted blue population at the bright end. By comparing
    subsamples of clusters that have a match between the two catalogues to
    those without matches, we discover that the CLF shape is mainly caused
    by the different cluster centroiding. However, the average weak lensing
    halo mass between the matched and non-matched clusters are
    consistent with each other in either cluster sample for halos with 
    $\lambda>30$ 
    ($\mhwl{>}1.5\times10^{14}\hmsol$). Since the colour
    preferences of the two cluster finders are almost orthogonal, such a
    consistency indicates that the scatter in the mass-richness relation of
    either cluster sample is close to random. Therefore, while the choice
    of how optical clusters are identified impacts the satellite content,
    our result suggests that it should not introduce strong systematic
    biases in cluster cosmology, except for the $\lambda<30$ regime.
\end{abstract}
\begin{keywords} galaxies: evolution --- galaxies: formation --- galaxies: abundances --- galaxies: statistics --- cosmology: large-scale structure of Universe --- gravitational lensing: weak
\end{keywords}

\section{Introduction}
\label{sec:intro}

Galaxy clusters are one of the most sensitive probes of dark energy and
cosmic growth~\citep[see section 6 of][for a comprehensive
review]{Weinberg2013}, but cluster cosmology in the optical is currently in
a state of minor crisis. The most recent cosmological constraints using
galaxy clusters~\citep[e.g.,][]{abb20,cos21, par21} favored lower
$\Omega_m$ and higher $\sigma_8$ compared against the constraints inferred
from the cosmic microwave background~\citep{pla20}, large-scale structures
traced by galaxies~\citep{ala17, beu11, abb18}, and clusters detected with
Sunyaev-Zeldovich~\citep{boc19} and X-ray~\citep{mantz15, chi22}
observations. To resolve this discrepancy, much work has been done to
determine which systematic uncertainties may be responsible. For instance,
projection effects are believed to boost the richness~(i.e., the number of
galaxies above some luminosity threshold) and the large-scale weak
lensing~(WL) signal of optical clusters~\citep{zu17,bus17}, and therefore
potentially bias cosmological
constraints~\citep[e.g.,][]{abb20,sun20,myl21,wu22, zen22}.

Although the projection effects in optical cluster cosmology are
problematic, they may mask a larger overall issue. The methods currently
used to optically identify galaxy clusters are imperfect and have
existing biases that must be taken into account. For the past twenty years,
the two primary methods used to identify optical clusters from photometric
surveys are the red sequence-based cluster finder, which uses overdensities
of ``red and dead'' galaxies on the sky to select cluster
candidates~\citep[e.g.,][]{gla00,gla05,mil05,ryk12,ryk14,ryk16} and the
halo-based method~\citep{yan05, yan07, yan21}, which instead relies on the
overdensities of all galaxies above some magnitude threshold, including the
blue galaxies that are often excluded by the red sequence~\citep[also
see][]{wen12,tin21b, zou21, tin22}. Although these methods greatly differ,
both have been used to successfully identify large samples of cluster
candidates with reasonably high completeness and purity.

Both methods of identifying galaxy clusters unavoidably have their own
preferences for the satellite content.
Identifying clusters using the red sequence works well
at $z{<}1$ where clusters have long been observationally dominated by
strong red sequences~\citep{gla00}. Under the assumption that the red
sequence forms as a result of halo quenching~(i.e., the probability of a
galaxy being quiescent depends only on halo mass~\citep{bir03, zu16}) red
sequence-based cluster finders should be efficient at sifting out the most
massive systems without introducing any biases due to the colour selection.
However, red-sequence clusters may be more strongly impacted by halo
assembly bias~\citep[e.g.,][]{gao05,gao07, jin07,dal08}, as these
overdensities of quiescent galaxies may be older clusters with higher
concentrations that more efficiently formed at early times~\citep{zu21,
zu22}. Additionally, this method does not account for the blue galaxies
that observations have found exist abundantly within the cluster
environment, which may impact the observational properties of the cluster.

Initially developed with the spectroscopic galaxies in mind~\citep{yan05,
yan07}, the halo-based method has been shown to work across a wide redshift
range and have a high completeness of clusters even when applied to
imaging data~\citep[][]{yan21}. This method evaluates the membership
probabilities for all galaxies regardless of colour, and relies on the
total luminosity of all member galaxies as a proxy of halo mass, which may
result in weighting blue galaxies more favorably than the red ones. The
primary bias for this method results from photometric-redshift estimates,
which are poorer for blue galaxies, leading to contamination in the
membership due to stronger projection effects.
In the current work, we exploit the fact that
the detection preferences associated with the red-sequence and halo-based
finders are roughly orthogonal, and investigate the differences not just
between the two cluster catalogues, but also between clusters that are
co-detected by both finders and those detected only once.

One primary statistical measurement used to characterize the galaxy content
within clusters is the conditional luminosity
function~\citep[CLF;][]{yan03}, defined as the luminosity function of all
the cluster member galaxies at fixed halo mass. Traditionally, the CLFs of
photometric clusters are measured by directly counting the number of member
galaxy candidates identified by the cluster finder, sometimes weighted by
the inferred membership probabilities~\citep[e.g.,][]{to20}. However, such
a direct method is incapable of recovering the true underlying CLFs of the
haloes, primarily due to biases in the colour selection and background
contamination caused by projection effects~\citep{zu17}. Instead of using
the membership provided by cluster catalogues, one could take advantage of
the excellent photometric redshifts of clusters, and measure the CLFs by
cross-correlating clusters with galaxies from imaging
data~\citep{han09, lan16, men22}. This approach accounts for the CLF
contribution of all galaxies regardless of colour, and the contamination
from interloper galaxies from distant, uncorrelated structures can be
statistically removed by subtracting the measurement around random points
on the sky. Moreover, the contribution of galaxies that are outside the
3D radius of the cluster but in a physically correlated structure along the
line-of-sight~(e.g., in the filaments that connect to the cluster) can be
modelled theoretically~\citep{lan16}. As a result, instead of measuring
the CLF properties of individual clusters, this cross-correlation or
``stacking'' approach measures the parameters associated with ensembles of
clusters that are stacked together based on a given richness. In this
work, we follow this stacking methodology to explore the different CLFs of
clusters detected by red-sequence and halo-based cluster finders.

Robust measurements of CLFs depend critically on the correct identification
of central galaxies, which is unfortunately challenging to
achieve for current optical cluster finders using just the imaging data
\citep[e.g.,][]{joh07,geo12,hol19}. Miscentring would result in a partial
overlap between the measurement aperture and the true cluster region,
leading to an underestimation of the CLF amplitudes and, more problematic,
biases in the shape of the CLFs due to the clumpiness of satellite
distribution within clusters. This issue can be statistically remedied in
cluster weak lensing studies by introducing a miscentring kernel
that can be calibrated against the X-ray centres \citep{zha19} and
convolved with a theoretical dark matter density profile \citep{nav97}.
However, the correct underlying CLFs cannot be reconstructed with the same
procedure, primarily because the true CLF model is unknown~(especially at
the faint end). In this work, for the first time we investigate the impact
of miscentring on the measured shape of the CLF, by examining the CLFs of
the same clusters but centred differently by the two cluster finders.

Discrepancies in the galaxy content between clusters detected with
different cluster finders do not necessarily introduce significant biases
in cosmological analyses. For a volume-complete cluster sample above some
true halo mass $M_h$, the red sequence-based and halo-based cluster finders
may both detect all the clusters but assign them richnesses with differing
amounts of logarithmic scatter at fixed halo mass. In the ideal scenario,
such a scatter in the mass-richness relation could be purely statistical in
nature, and thus can be self-calibrated when constraining cosmology with
clusters~\citep{roz10, tin12, zu14}. However, if the cluster finder
preferentially assigns higher richnesses to haloes with a particular CLF,
such as having a higher red fraction due to older formation time, or higher
blue fraction due to stronger projection effects within dense environments,
the scatter is no longer random and would systematically bias the
constraint on cosmological parameters. In this work, we take advantage of
the orthogonal CLF preferences between the red-sequence and halo-based
finders, and divide each catalogue into two subsamples based on whether the
clusters are also detected by the other cluster finder. Therefore, by
examining whether there exists any significant discrepancy in the average
halo mass measured from weak lensing between the two subsamples, we can
better understand the nature of the scatter in the mass-richness relation
for different cluster finders.

The structure of the paper is as follows. In Section~\ref{sec:data}, we
describe the photometric galaxy catalogue and cluster samples selected for
this analysis. In Section~\ref{sec:CLF}, we describe how the CLFs are
measured in both samples and compare the galaxy content revealed by their
CLFs. In Section~\ref{sec:WL} we describe the cluster weak lensing
measurements and compare the weak lensing halo masses between the matched
vs. non-matched clusters within each sample. We conclude the paper and look
to the future in Section~\ref{sec:conclusions}. Throughout this analysis,
we assume a flat $\Lambda$CDM characterized by \citet{pla20} cosmology with
$\Omega_{M}$=0.315 and $h$=0.674. Additionally, we note that unless
explicitly stated, we are using $\log\,M_h$ for the logarithm with base 10
of halo mass in unit of $\hmsol$, defined by the halo radius within which
the mean halo density is $\Delta{=}200$ times the mean density of the
Universe. Finally, all distances measured are co-moving distances in units
of $\hmpc$.

\section{Data}
\label{sec:data}

\subsection{The SDSS Photometric Galaxy Sample}
\label{subsec:SDSS}

In this analysis, we measure the CLF by statistically counting the galaxies
from a photometric catalogue constructed using Sloan Digital Sky Survey
Data Release 16~\citep[SDSS DR16;][]{yor00,ahu20}. We select all galaxies
with clean photometry~(\texttt{CLEAN}=1) with an SDSS r-band Galactic
extinction-corrected model-magnitude greater than
21~(\texttt{modelmag\_r}$-$\texttt{extinction\_r}\texttt{<21.0}) for
completeness. In total, this yields a sample of about 47 million galaxies
within a sky coverage of approximately 8500 deg$^2$.

For measuring the CLF contribution from any given cluster at redshift, $z$, we
estimate the absolute magnitude M$_{r}$ for each galaxy assuming it is at
the redshift of that cluster as $M_{r}{=}m_r{-}\mathrm{DM}(z)$, where $m_r$
is the apparent magnitude after being corrected for Galactic extinction and
DM($z$) is the distance modulus at the redshift of the cluster. Interloper
galaxies that are not physically associated or correlated with the cluster,
hence assigned the wrong redshift, will be statistically removed by the
background subtraction~(as discussed in
Section~\ref{subsec:CLF-Measurement}).

We note that because our goal for this analysis is a comparison of the CLF
measured in clusters identified in the \citeauthor{ryk14} and
\citeauthor{yan21} catalogues within a relatively narrow redshift
range~($0.1{<}z{<}0.34$), we choose not to account for the K-correction,
which systematically shifts the CLFs by the same amount for both
measurements. Moreover, the K-correction to $z=0.1$ is of the order of
$\sim$0.1 magnitudes across our redshift range and given our average bin
size (0.4 magnitudes) when measuring the CLF would only have a minimal
impact on our presented results.

\begin{figure}
    \centering \includegraphics[width=0.48\textwidth]{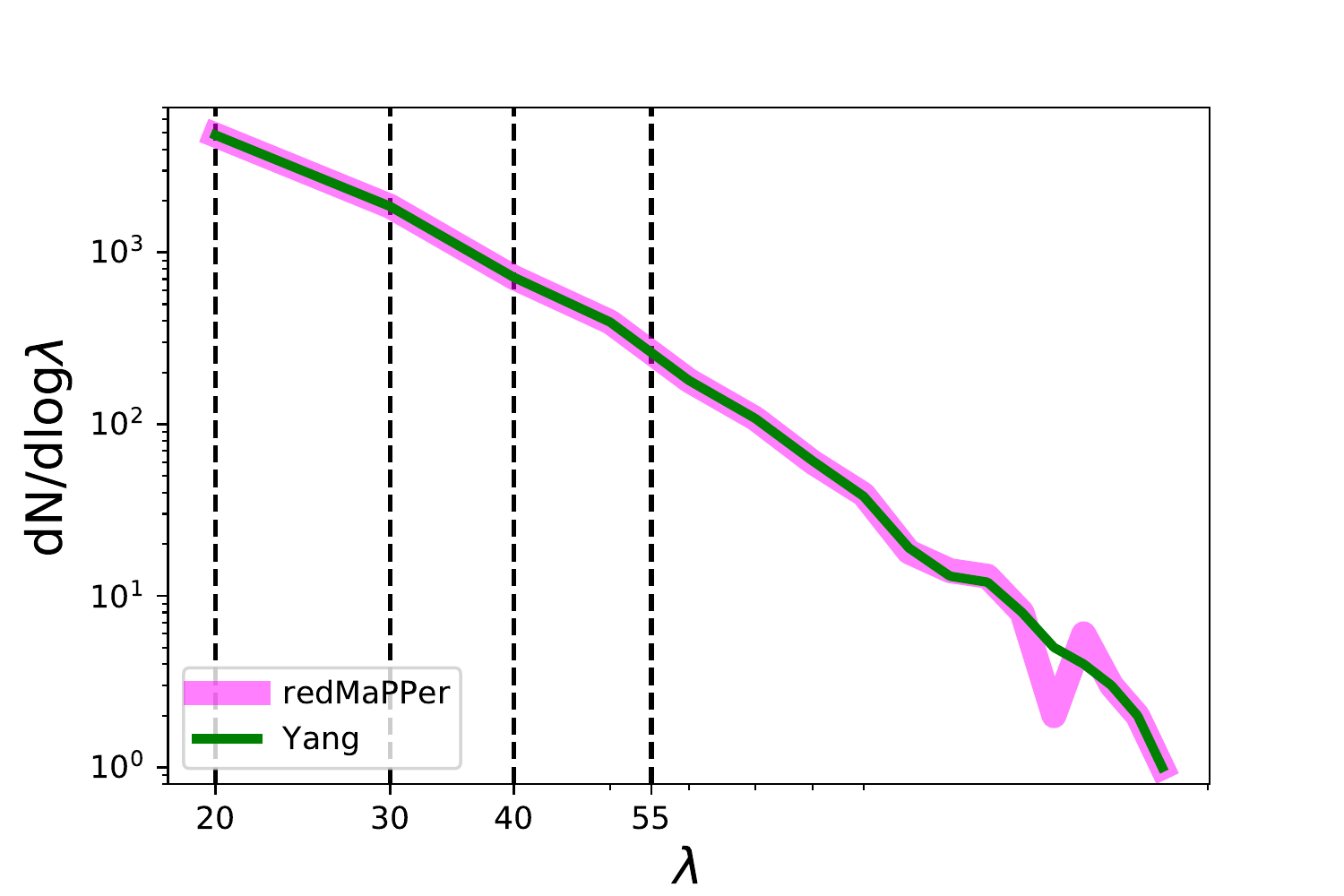}
    \caption{The richness distributions of
    the \yang{}~(thin green curve) and \redmapper{}~(thick magenta curve)
    samples. The richness estimates of \yang{} clusters are measured by
    counting the total number of member galaxies brighter than $M_r{=}{-}20.4$
    and enclosed within an $1\hmpc$ aperture, and then abundance-matched to
    the richness distribution of \redmapper{} clusters, hence the
    consistency between the two curves. Vertical dashed lines indicate the
    edges of the four richness bins adopted in our analysis.}
    \label{fig:AM}
\end{figure}

\subsection{Red Sequenced-based Cluster Sample: \redmapper{}}
\label{subsec:RM}

The red sequenced-based cluster catalogue we employ is the SDSS
\redmapper{} v6.3\footnote{https://risa.stanford.edu/redmapper/} cluster
catalogue~\citep{ryk14} derived by applying a red-sequence-based
photometric cluster finding algorithm to the SDSS DR8
imaging~\citep{aih11}. Briefly, \redmapper{} iteratively self-trains a
model of red-sequence galaxies calibrated by an input spectroscopic galaxy
sample, and then attempts to grow a galaxy cluster centred on every
photometric galaxy. Once a galaxy cluster has been identified by the
matched-filters, the algorithm iteratively solves for a photometric
redshift based on the calibrated red-sequence model, and recentres the
clusters about the best candidates for central galaxies.

Therefore, each \redmapper{} cluster is a conglomerate of red-sequence
galaxies on the sky, with each galaxy assigned a membership probability
$\pmem$ and a probability of being the central $\pcen$. For each cluster, the
richness $\lamrm$ is computed by summing the $\pmem$ of all member galaxy
candidates, and roughly corresponds to the number of red-sequence satellite
galaxies brighter than $0.2\,L_*$ within an aperture of
${\sim}1\,\hmpc$~(with a weak dependence on $\lamrm$). At
$\lamrm{\geq}20$, the SDSS \redmapper{} cluster catalogue is approximately
volume-complete up to $z{\simeq}0.34$ \citep{gro17}, with cluster
photometric redshift uncertainties as small as
$\delta(z)=0.006/(1+z)$~\citep{ryk14, roz15}.

We use the $\lambda{\ge}20$ sample between $z{=}0.1$ and $0.34$. For this
analysis, our initial sample of \redmapper{} clusters consists of the 9226
clusters between the redshift range $0.10<z<0.34$. For each cluster we
identify the most likely central galaxy candidate provided by
\redmapper{} and treat its location as the centre for the CLF measurement.

\subsection{Halo-based Cluster Sample of Yang et al. (2021) }
\label{subsec:Yang}

The halo-based clusters used in this analysis are detected by the
\citet{yan21}\footnote{https://gax.sjtu.edu.cn/data/DESI.html} cluster
finder~(hereafter referred to as the \yang{} clusters). This cluster finder
is an extension of the spectroscopic group finder previously developed by
\citet{yan05,yan07} and has been updated to incorporate both photometric
and spectroscopic information. The photometric data used to construct the
\citet{yan21} catalogue comes from the DESI Legacy Imaging Survey
DR9~\citep{dey19}, which covers the redshift range $0.0<z\le1.0$.

We briefly summarise the halo-based methodology used in the \yang{} cluster
finder below and refer interested readers to \citet{yan21} for the
technical details. The halo-based algorithm attempts to build groups or
clusters centred on every galaxy within the legacy imaging data by
assigning initial membership to galaxies using a friends-of-friends
approach assuming some small linking length~\citep{yan07}. Using the total
luminosity within these initial overdensities, the mass-to-light ratio of
each cluster is determined via abundance matching, which allows each
tentative cluster to be assigned a halo mass. Cluster membership is then
determined assuming the galaxy distribution follows the same phase-space
distribution as the dark matter particles. This requirement selects all
member galaxies above a background level within $R_{180}$ from the
luminosity-weighted cluster centre. This process is then repeated
iteratively (3-4 times) until the mass-to-light ratios converge. Using mock
galaxy samples, \citet{yan21} demonstrated that this adaptive halo-based
algorithm can reliably detect systems with mass above a few
${\times}10^{13}\hmsol$ with
a purity higher than 90\%.

\begin{figure*}
    \centering
    \includegraphics[width=0.8\textwidth]{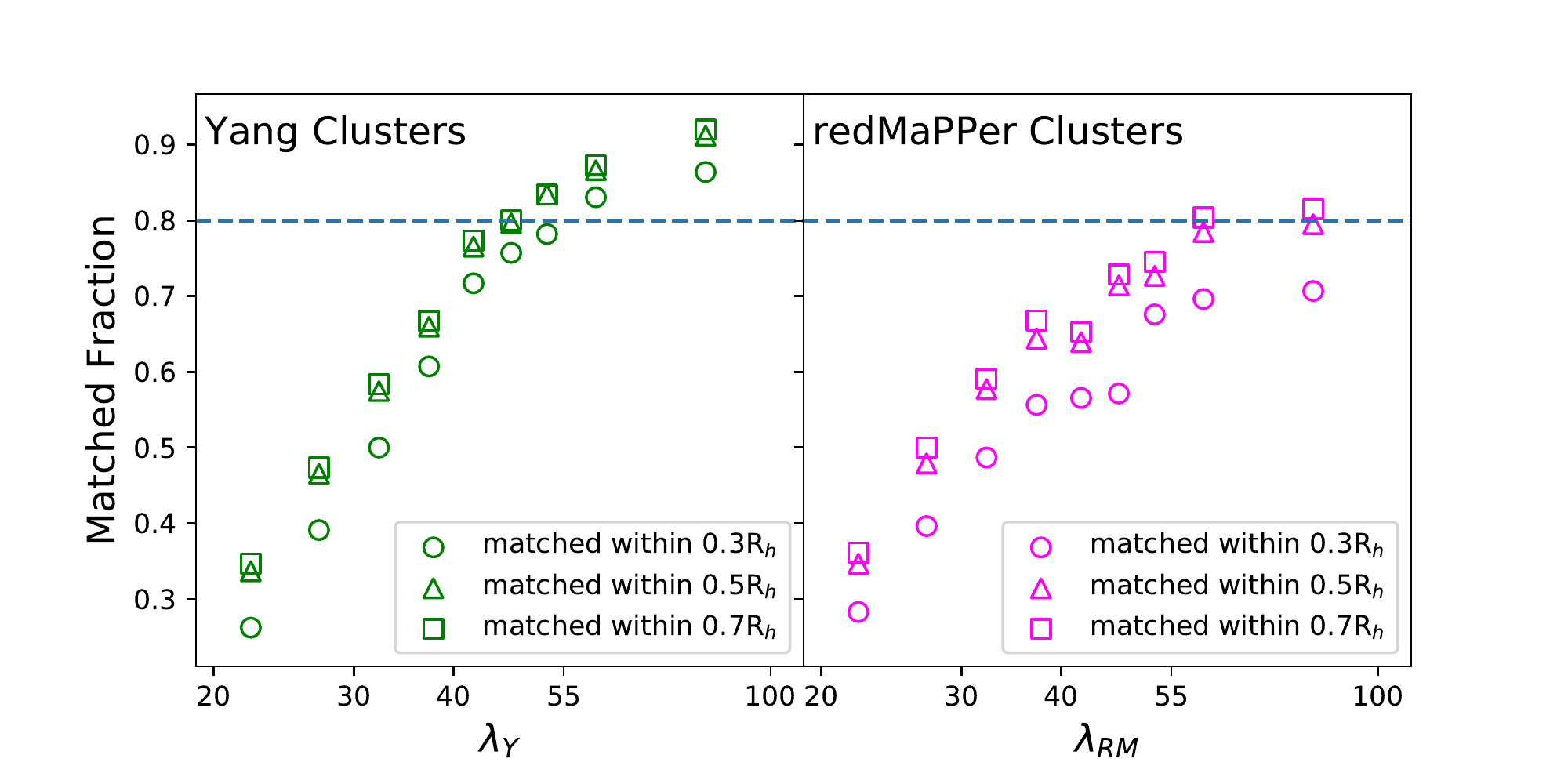}
    \caption{The matched fraction as a function of richness for the
    \yang{}~(left panel) and \redmapper{}~(right panel) cluster samples,
    using three different search radii of 0.3$R_h$~(circles),
    0.5$R_h$~(triangles), and 0.7$R_h$~(squares). We adopt the
    cross-matching results using the 0.7$R_h$ search radius as our fiducial
    split between matched vs. non-matched clusters in the paper.}
\label{fig:overlap}
\end{figure*}

The \yang{} catalogue provides an abundance-matched halo mass, $\mhy$,
defined with $\Delta{=}180$ instead of $200$ and based on the total
luminosity of galaxies in the cluster~(with DESI z-band apparent magnitude
${z_{\rm DESI}}$<21.0 within $R_{180}$). \citet{yan21} showed that the
scatter about the $\mhy$-$M_h$ relation is about 0.2 dex above
$10^{14}\hmsol$, similar to that in \redmapper{}. For centring,
instead of computing $p_{\mathrm{cen}}$ as done in \redmapper{}, the
\yang{} catalogue adopts the luminosity-weighted centre as the fiducial
centre of a cluster. For this analysis, we do not cover the entire
redshift range of the \yang{} catalogue, but instead use data only from the
redshift range over which \redmapper{} is complete as well as only those
clusters with $\log\mhy{\ge}14.0$. These cuts limit the \yang{} sample to
29014 clusters. We also note that although this catalogue was developed
using the DESI imaging, we measure the CLF for \yang{} clusters using the
same SDSS photometric data as for \redmapper{}.

\subsection{\redmapper{} and \yang{} Clusters within the Same Volume}
\label{subsec:analysis}

The purpose of our CLF analysis is to compare the galaxy populations within
haloes identified by either \redmapper{} or the \yang{} cluster finder.
However, these two cluster catalogues are identified using different galaxy
surveys and cover different spatial and redshift regions. To account for
this, we only select clusters in the overlapping footprint between the two
samples and within the redshift range $0.1<z<0.34$ where both samples
are reasonably complete at the high mass end. The DESI legacy imaging
footprint is larger than the region covered by SDSS, so much of the data
removed is from the \yang{} catalogue. In total, the number of
\redmapper{} clusters was reduced to 8770, a reduction of less than 5\%.
In contrast, the number of \yang{} clusters was reduced to 15063, a
reduction of 48.1\%. Moreover, to accurately compute the covering
fraction of any regions on the sky, we require that each cluster be within
the SDSS BOSS spectroscopic footprint, for which convenient window and mask
polygon files\footnote{https://data.sdss.org//sas/dr12/boss/lss/} are
available. Removing clusters outside the BOSS window~(minus the bright
star masks) function slightly reduces the number of available clusters in
each footprint. The \redmapper{} cluster sample then consists of 8300
clusters and the \yang{} sample consists of 14155 clusters, a reduction of
5.4\% and 6.0\%, respectively.

To facilitate the comparison between these two cluster samples, we need to
account for the difference in the aperture sizes adopted by the two cluster
finders when defining their respective halo mass proxy. In particular,
\redmapper{} provides a richness~($\lamrm$) measured within roughly
1$\hmpc$, while the $\mhy$ of \yang{} clusters are estimated within
$R_{180}$, which is usually larger than 1$\hmpc$. Therefore, for each
\yang{} cluster, we redefine a new richness $\lamyprime$ by
measuring the total
number of member galaxies~(with equal weights) with SDSS r-band absolute
magnitudes $M_{r}{<}{-}20.4$, enclosed within a fixed $1\hmpc$ aperture
centred on the luminosity-weighted centre. The member galaxies are directly
adopted from the \yang{} membership catalogue, which is roughly
volume-complete above $M_{r}{>}-20.4$ between $0.1{<}z{<}0.34$.

To further bring the two richness measurements~($\lamyprime$ vs. $\lamrm$)
to the same scale, we perform a standard abundance matching~(assuming zero
scatter) to ensure that the richness distribution of the top 8300 \yang{}
clusters~(ranked by $\lamyprime$) is the same as the $\lamrm$ distribution
of the 8300 \redmapper{} clusters. After abundance matching, we reassign a
richness value $\lamy$ to each of the top 8300 \yang{} clusters, and adopt
$\lamy$ for the \yang{} clusters for the remainder of the paper.
Figure~\ref{fig:AM} shows the richness distributions of the
\redmapper{}~(thick magenta curves) and \yang{}~(thin green curves)
clusters within the same volume, using the $\lamrm$ and $\lamy$ estimates,
respectively. Going forward, unless otherwise noted we will refer to both
richnesses as $\lambda$.

Using the two abundance matched samples, we follow the approach used in
\citet{lan16} and only select clusters into our final analysis samples
with a covering
fraction within 1$\hmpc$ from the cluster centre higher than 95\%. This cut is
needed to remove clusters located at the edges of the survey as well as
those near regions that are masked by bright foreground stars. This
criteria reduces the number of clusters in our final sample to 7582 for
\redmapper{} and 7562 for \yang{}, reductions of 8.7\% and 8.9\%,
respectively.

Finally, for the CLF and weak lensing analyses we divide each cluster
sample into four subsamples based on $\lambda$, indicated by the vertical
dashed lines in Figure~\ref{fig:AM} --- the four richness bins are
$\lambda=[20,\,30]$, $[30,\,40]$, $[40,\,55]$, and $[55,\,200]$,
respectively.

\begin{figure*}
    \centering
    \includegraphics[width=0.98\textwidth]{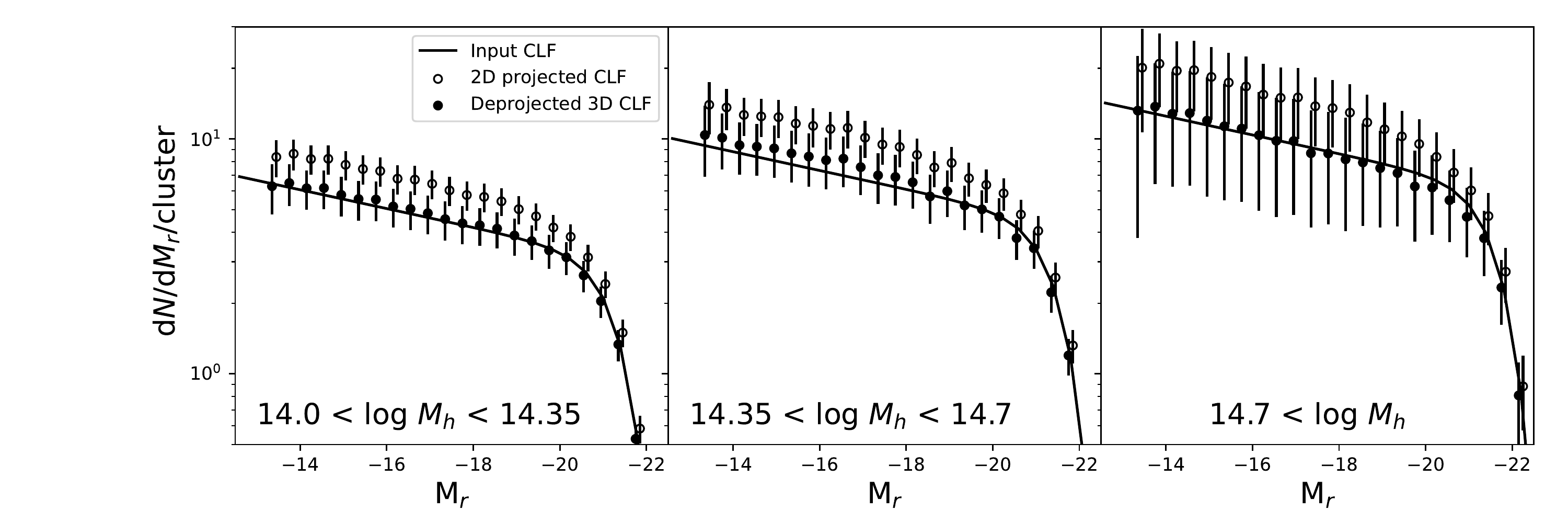}
    \caption{Validation of the CLF deprojection method using mock
    photometric galaxies in three different halo mass bins. In each panel, the
    black solid line is the input CLF from~\citet{yan08}, open circles are
    the 2D projected CLF measured from the stacking method, and filled
    circles are the 3D deprojected CLF from the 2D results, using the
    magnitude-dependent correction factor at each halo mass bin
    $f_{\mathrm{corr}}(M_r)$. There is excellent agreement between the
    deprojected 3D CLF and the input CLF.} \label{fig:deprojection}
\end{figure*}

\subsubsection{Cross-Matching Cluster Samples }
\label{subsubsec:Overlap}

As discussed in Section~\ref{sec:intro}, we also investigate the properties
between the clusters that are co-detected by both cluster finders and those
detected in one but not in the other. To cross-match between the two
catalogues, for a cluster from one catalogue, we identify from the other
catalogue any cluster detected within some projected 2D radial aperture and
with a redshift within 3$\sigma_z$ as a match, where $\sigma_z$ is the
uncertainty on the photometric redshift estimate. For \redmapper{} this is
$\delta(z)/(1+z)$ and $\approx$0.006 at $z=0.1$ and $\approx$0.015 at $z=0.3$
\citep{ryk14, roz15}; For \yang{} clusters this is 
$\sigma_{z}\approx0.008$~\citep{yan21}. For this cross-matching, 
we only use the two final
samples, so a cluster classified as ``non-matched'' might have a
matched counterpart with $\lambda<20$ in the other cluster catalogue. Were
we to use the entire samples~(down to much lower $\lambda$), as
well as extend out to arbitrarily larger radii, the matched fraction would
increase significantly, thereby making the comparison less useful.

As shown in Figure~\ref{fig:overlap}, we use three different search radii,
0.3R$_{h}$, 0.5R$_{h}$ and 0.7R$_{h}$. For the sake of simplicity we use
$R_h$ that corresponds to the original apertures adopted by the cluster
finders. We see that for the most
massive clusters ($\lambda{\simeq}55$), the matched fraction
is greater than 80\% for a search radius greater than 0.5R$_{h}$.
Unsurprisingly, the matched fraction decreases with decreasing richness,
as the matches are increasingly more
likely to be found in the $\lambda<20$ systems. Besides, the completeness
and purity of both cluster catalogues also drop rapidly with richness. We
note that a more complete analysis of the comparison between the clusters
and groups identified by \redmapper{} and the \yang{} cluster finder is
needed, but beyond the scope of this analysis. For the analysis in this
paper, we adopt 0.7R$_{h}$ as the fiducial search radius for our
matched vs. non-matched split.

\section{The Conditional Luminosity Functions}
\label{sec:CLF}

\subsection{Stacked 2D CLF Measurement}
\label{subsec:2dclf}

To compare the satellite populations within clusters from both catalogues,
we measure the CLF for clusters within each $\lambda$ bin. The standard
approach~\citep[e.g.,][]{han09,lan16} is to separate the central and
satellite components of the CLFs. However, because the \yang{} clusters are
not centered on the central galaxy and the central is included when
measuring the CLF within 1$\hmpc$ of the luminosity weighted center, we
include the central galaxy component in the measurements of all CLFs in
this analysis. Despite this inclusion of centrals in the CLFs, we only
focus on the magnitude range dominated by the satellite galaxies and refer
to the member galaxies at those magnitudes simply as ``satellites'' during
our CLF comparison. In an upcoming analysis, we plan to build on this work
and use the central galaxy to investigate the impact of the magnitude gap,
a tracer of BCG hierarchical growth \citep{gol18}, on the CLF.

Unlike some previous analyses \citep[e.g.,][]{to20}, we do not
directly use either the \redmapper{} or the \yang{} membership candidates
to construct the CLF (though we present that direct result as well).
Instead, we closely follow the stacking procedures of \citet{lan16} to
measure the CLF of both cluster catalogues using the SDSS photometric
galaxies. Using the centroid positions identified in those catalogues, for
each galaxy cluster, we identify all galaxies within a projected distance
of 1$\hmpc$. We use a constant radial aperture to mitigate the impact of
uncertainties in the halo mass estimates. We then convert their apparent
magnitudes into absolute magnitudes assuming the cluster redshift. However,
as previously noted because this is a photometric measurement, we must
account for contamination from interloper galaxies. For each cluster, we
assign the redshift and halo mass to ten random points within the survey
footprint and use the same 1$\hmpc$ aperture to determine the background
contribution. By subtracting off the background from the measurement
centred on the cluster centre, we obtain the 2D CLF
measurement for each individual cluster.

We next employ the BOSS window function~(with bright stars masked out) to
estimate the covering fraction of each cluster region. More specifically,
we determine what fraction of the sky within a 1$\hmpc$ aperture is not
masked as a result of bright stars or edge effects. The covering fractions
are generally high as our final analysis sample includes only those clusters
with covering fraction greater than 95\%.
We multiply the measured CLF by the inverse of this fraction to correct for
masking. Additionally, when stacking the measurements, we use the limiting
magnitude of the SDSS observations ($m_r{=}21$) to calculate the absolute
magnitude limit for each cluster and correct for this difference across our
redshift range $0.1{<}z{<}0.34$. We then calculate the stacked 2D CLF by
measuring the mean 2D CLF within 1$\hmpc$ of the cluster centre~(with the
mean random background contribution subtracted) for each of the four
richness bins described in Section~\ref{subsec:analysis}.

At this point, we are left with a 2D CLF, which is representative of the
average galaxy population projected within a 1$\hmpc$ 2D aperture about the
cluster centre. However, our goal is to estimate the 3D isotropic CLF
within a 1$\hmpc$ radius, for which we need to further remove the CLF
contribution from galaxies in the correlated structures but physically
outside the 1$\hmpc$ radius.

\subsection{Deprojecting 2D CLF to 3D: Mock Test using {\it N}-body
Simulation}
\label{subsec:Cal}

\begin{figure*}
    \centering
    \includegraphics[width=0.98\textwidth]{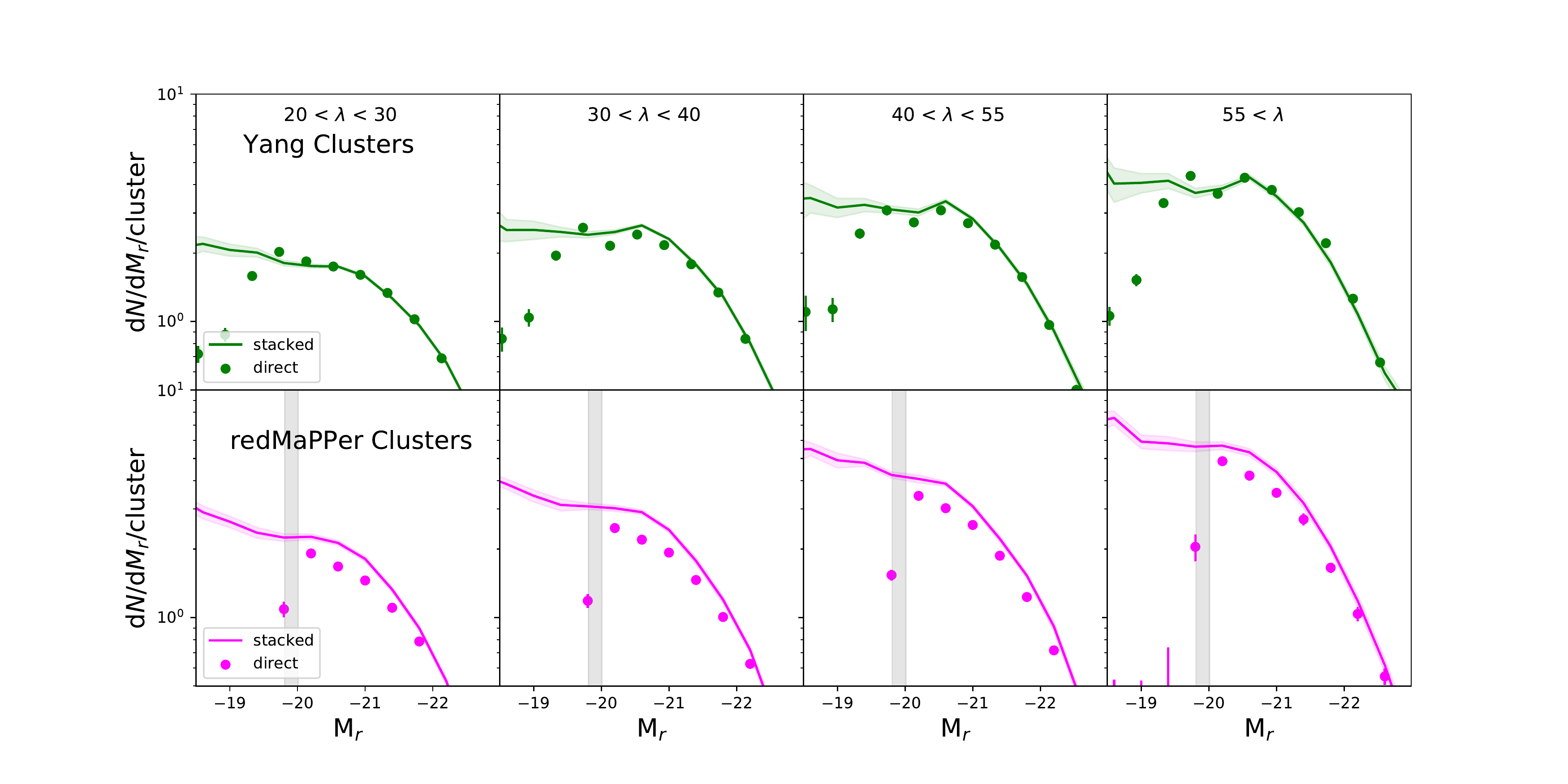}
    \caption{For four bins in $\lambda$, we compare the CLF measured directly
    from the membership catalogues~(circles with errorbars) to those
    measured from cross-correlating with the photometric galaxies~(curves
    with uncertainty bands) in both the \yang{}~(top row) and
    \redmapper{}~(bottom row) catalogues. In the lower panels, we
    denote $0.2 L_*$ as the gray band.}  \label{fig:members}
\end{figure*}

We compute the correction factor to deproject the CLF measurements from 2D
to 3D by following the approach outlined in \citet{lan16}. Although such a
deprojection approach has been applied to real data by \citet{lan16},
we verify our capability of accurately measuring the average CLF per
cluster within a 3D radius using a sample of simulated clusters and a mock
catalogue of photometric galaxies. In particular, we determine whether any
additional calibrations are needed to accurately remove the contribution
from correlated structures to the CLF when deprojecting the 2D measurements
to 3D.

Using the ELUCID {\it N}-body constrained cosmological
simulation~\citep{wan16}, we construct the mock photometric galaxy
catalogue by populating dark matter haloes according to the halo
mass-dependent CLF parameters measured by \citet{yan08}. For a continuous
coverage of the ELUCID halo mass function with galaxies down to $m_r{=}21$,
we interpolate the CLF parameters across the three most massive bins of
\citet{yan08} and extrapolate the CLFs of \citet{yan08} to fainter
magnitudes assuming fixed faint-end slopes. We further assume
the galaxy distribution
follows an isotropic NFW profile inside the dark matter haloes with the same
concentration and velocity dispersion as the dark matter particles.
Following \citet{sal22}, we populate the halos within the SDSS main sample
volume inside the ELUCID constrained simulation.

Following the approach outlined in the Appendix B2 of \citet{lan16}, we
calculate a correction factor $f_{\mathrm{corr}}$ to account for the
discrepancy that exists between the CLF projected within a 2D aperture and
that enclosed within a 3D halo radius, which results from galaxies
associated with the correlated structure outside the clusters. We briefly
describe the deprojection method below and refer interested readers to
\citet{lan16} for the analytic derivations. To determine this correction
factor at any given magnitude for each halo mass/richness bin,
$f_{\mathrm{corr}}(M_r)$, we measure the normalized surface
number density of galaxies at a given magnitude between 0.1R$_{200}$ and
2.5R$_{200}$ and fit the power-law slope, $\gamma$, to this profile. We
then use equation B7 from \citet{lan16} to convert the power-law slope
$\gamma$ to the correction factor $f_{\mathrm{corr}}$. We note that this
process differs from \citet{lan16} in that we adopt a magnitude-dependent
$\gamma$ to remove the contribution from the correlated structure instead
of measuring the median value of $\gamma$ within each richness bin.

As shown in Figure~\ref{fig:deprojection}, the 2D CLF overpredicts the
number of galaxies associated with each cluster. However, after applying
the correction factors to the 2D CLF, our deprojected 3D CLF is in
excellent agreement with the theoretical input from \citet{yan08}. Thus,
the magnitude-dependent correction factors are needed to accurately
reproduce the input CLF. Therefore, in our observational analysis, we
measure the surface density profile out to 2.5R$_{200}$ at fixed $M_r$ for
each cluster and measure the correction factor as a function of $M_r$ for
clusters of each richness bin.

\subsection{Stacked 3D CLFs: Comparison with Direct CLFs}
\label{subsec:CLF-Measurement}

After testing the stacking and deprojection method with mock galaxy
samples, we are now ready to apply the method to the \redmapper{} and
\yang{} cluster samples using the SDSS photometric catalogue. We apply the
correction factor $f_{\mathrm{corr}}(M_r | \lambda)$ to each 2D CLF to
remove the effect of galaxies that are outside of our 1$\hmpc$ radius when
measured in 3D, thereby converting our previously measured 2D CLFs to
the deprojected 3D CLFs.

Although both the \redmapper{} and \yang{} cluster catalogues provide
membership catalogues to identify the galaxies associated with each
cluster, the uncertainties associated with the membership probabilities are
generally large, with significant biases due to the redshift uncertainties
and projection effects~\citep{zu17}. Therefore, CLFs constructed directly
from those membership catalogues~(hereafter referred to as ``direct'' CLFs)
are expected to be incomplete. Figure~\ref{fig:members} presents the
results of the direct CLF when measured within 1$\hmpc$ for each membership
catalogue, compared to the 3D CLF measured from the stacking
method~(band). For the direct CLFs of \yang{} clusters, we apply a constant
shift of 0.07 magnitudes to convert the DESI magnitudes of the member
galaxies to SDSS r-band magnitudes $M_r$. We note that \redmapper{} also
provides a membership probability $\pmem$, which we use in our estimate of
the direct CLF.

As shown in Figure~\ref{fig:members}, we find very similar trends between
the direct and stacked CLFs for both the \yang{}~(top row) and
\redmapper{}~(bottom row) clusters at the bright end~($M_r{<}{-}21.5$),
indicating that both cluster finders are highly effective in identifying
the bright member galaxies. At the faint end, we note that \redmapper{}
uses a limiting magnitude of $0.2L_{*}$~(gray vertical bands) to determine
its membership catalogue. As a result, it is unsurprising that we see the
\redmapper{} direct CLF~(filled circles) drop off beyond that limit.
Meanwhile, the \yang{} catalogue imposes an apparent magnitude cut on
the membership galaxies, so we do not observe a steep drop-off in the
direct measurement like what is seen in \redmapper{}, and their direct and
stacked CLFs are in reasonable agreement until $M_r{\sim}{-}19.5$.
Interestingly, the direct CLFs of both \yang{} and \redmapper{} exhibit a
modest deficit compared to
their respective stacked CLFs at ${-}20{>}M_r{>}{-}21$, suggesting that
both cluster finders are relatively conservative in assigning memberships
to galaxies with intermediate luminosities between sub-$L_*$ and $L_*$.
However, we note that the deficit should be caused by the incompleteness of
different types of galaxies, (relatively redder galaxies in \yang{} and
bluer galaxies in \redmapper{}).

\subsection{CLF Comparison between the \yang{} and
\redmapper{} Clusters}
\label{subsec:CLF-results}

\subsubsection{CLF Amplitudes}
\label{subsubsec:normalisation}

\begin{figure}
    \centering \includegraphics[width=0.48\textwidth]{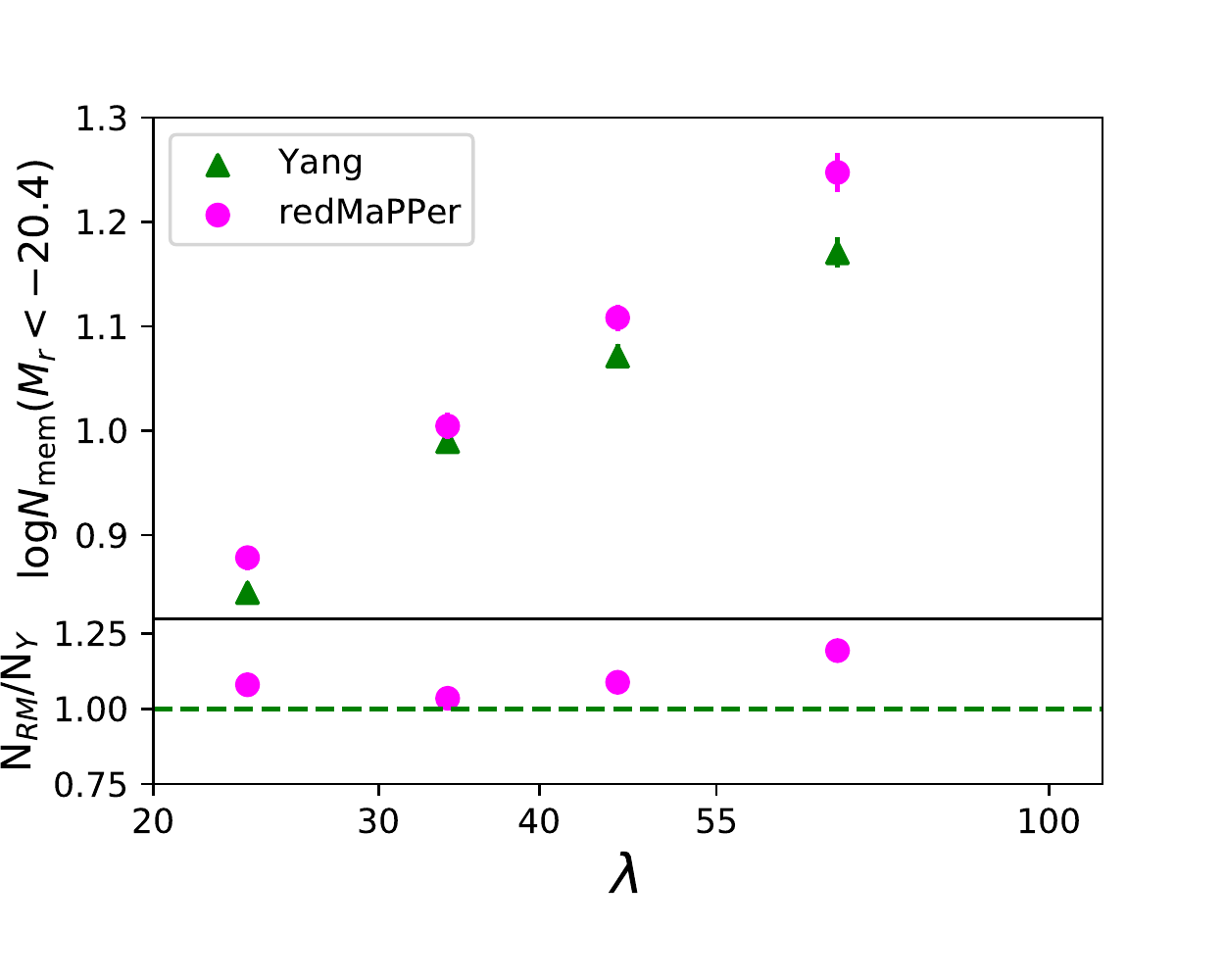}
    \caption{Comparison between the total number of member galaxies
    $\nsat$ between the \yang{}~(green triangles) and \redmapper{}~(magenta
    circles) clusters in four different richness bins. The satellite numbers are
    computed from integrating the stacked CLFs above $M_r{=}{-}20.4$. The
    lower panel shows the ratio of the number of members in the
    \redmapper{} clusters divided by that in the \yang{}
    clusters.}
    \label{fig:Numgal}
\end{figure}

We compute the total number of member galaxies brighter than $M_r{=}-20.4$,
$\nsat$, by integrating the stacked CLFs of both cluster samples at fixed
$\lambda$~(as shown in Figure~\ref{fig:members}) above ${-}20.4$.
Figure~\ref{fig:Numgal} compares the $\nsat$ between the \yang{}~(green
triangles) and \redmapper{}~(magenta circles) clusters in the four richness
bins, with the ratios $N_{\mathrm{RM}}/N_{\mathrm{Y}}$~(circles) shown in
the bottom panel. The number of member galaxies at the lower richness end
($\lambda$$\le$55) are in good agreement~(within $10\%$). In contrast, at
the higher richness end the \redmapper{} clusters contain $\sim$20\% more
member galaxies. By using a fixed $1\hmpc$ aperture, we eliminate the
impact of halo definitions. Therefore, the discrepancy is likely driven by
the combination of different centring and scatters in the mass-richness
relation between the two catalogues. In particular, it is likely that 1)
the centrals identified by \redmapper{} are located in denser regions than
the luminosity-weighted centre from the \yang{} catalogue; and 2) the new
$\lamy$ richness that we redefined for the \yang{} catalogue has a slightly
larger scatter~(at fixed halo mass) than $\lamrm$ and $\mhy$~(i.e., the
original halo mass proxy of the \yang{} catalogue) at the high mass end,
thereby finding intrinsically lower mass clusters at the same $\lambda$.
However, as discussed further in Section~\ref{sec:WL}, our cluster weak
lensing analysis suggests that the relatively larger scatter in our new
richness $\lamy$ is likely the more dominant cause, which is expected
because $\lamy$ is not optimized for minimising the scatter in the
mass-richness relation. Additionally, we note that when using the BCGs
identified by \yang{}~(not shown in this analysis), this $\nsat$
discrepancy is enlarged, suggesting that the luminosity-weighted centres
are indeed better choices for the CLF measurement.

\subsubsection{Dependence of the CLF Shape on Galaxy Colour and Cluster
Finders}
\label{subsubsec:shape}

\begin{figure*}
    \centering
    \includegraphics[width=0.98\textwidth]{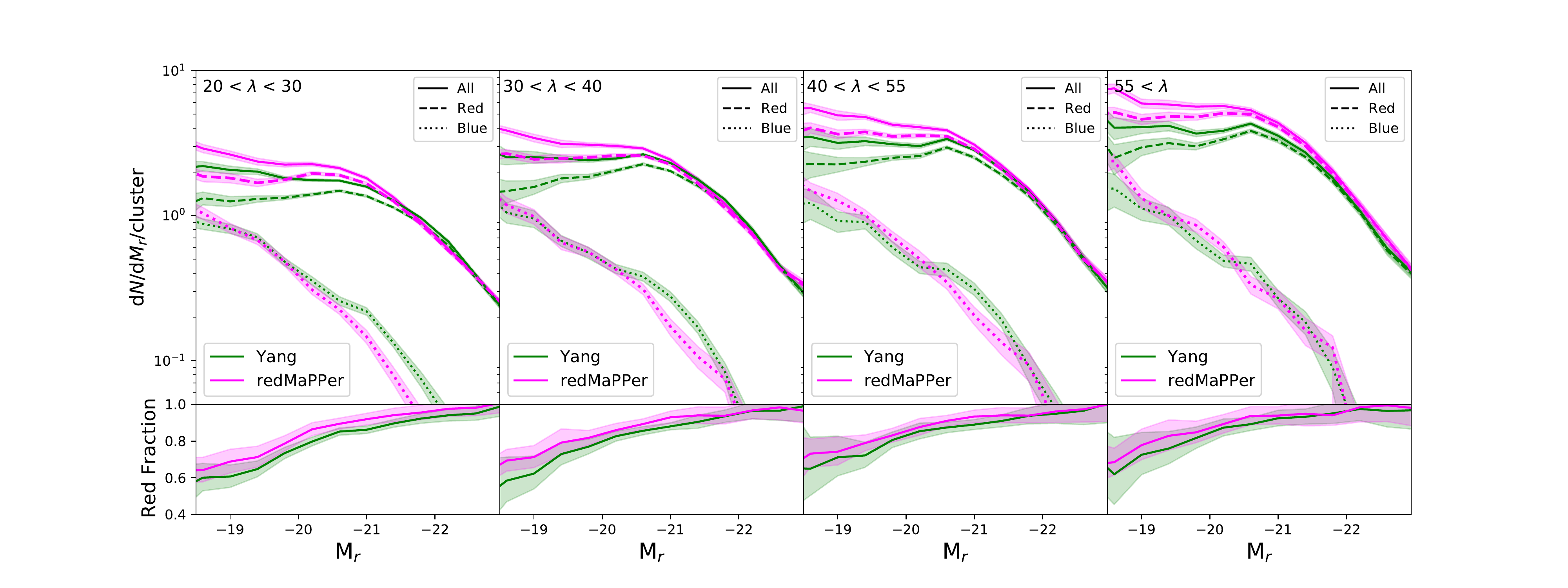}
    \caption{Comparison of the CLF shapes between the \yang{}~(green) and
    \redmapper~(magenta) clusters in four richness bins. In each upper
    panel, the total CLF~(solid curves) is decomposed into the
    contributions from red~(dashed curves) and blue~(dotted curves)
    galaxies. The lower panels show the red fractions as functions of the
    absolute r-band magnitude.}
    \label{fig:CLF_RM_Yang}
\end{figure*}

We now focus on the shapes of the stacked CLFs to compare the galaxy
content at different absolute magnitudes between the \redmapper{} and
\yang{} clusters.  Additionally, to examine the colour difference in the
galaxy content, we split our photometric galaxies by colour to measure the
stacked CLFs of the red and blue satellites separately. Unlike
\citet{lan16}, who used $u{-}r$ colour, or \citet{men22}, who used $g{-}z$
colour, we use the redshift-dependent \redmapper{} red-sequence defined in
$g{-}r$ to identify all red galaxies across the redshift range of interest.
This approach also accounts for the shallow slope of the red sequence on
the colour-magnitude diagram at any given redshift. We then identify
galaxies with $g{-}r$ colours 2$\sigma$ below the red sequence as blue and
all other galaxies as red.

Figure~\ref{fig:CLF_RM_Yang} compares the CLFs between the \redmapper{} and
\yang{} clusters using the two overall samples~(i.e., regardless of whether
a cluster in one sample has a matched counterpart in the other sample).  At
the bright end ($M_{r}<-20.4$), where the massive red galaxies dominate the
CLFs, the shapes are in excellent agreement. However, the faint-end slopes
of the overall CLF~(solid curves) differ, with
$\alpha_{\mathrm{Yang}}{\simeq}{-}0.90$~(green) and
$\alpha_{\mathrm{RM}}{\simeq}{-}1.1$~(magenta). This discrepancy appears to
be driven by the lower luminosity red galaxies (dashed curves), where
\redmapper{} clusters exhibit a relatively constant population, compared to
the slightly decreasing population in \yang{} clusters.  Meanwhile, we
detect a boost in the population of blue galaxies in the \yang{} clusters
at $M_{r}{<}{-}21$ compared to the \redmapper{} clusters, and the 
amplitude of the
boost declines with increasing richness, exhibiting little difference
between the two blue CLFs in the highest-$\lambda$ bin.  Both discrepancies
in the red and blue CLFs are consistent with our expectation that the
colour preferences of the red-sequence and halo-based cluster finders are
largely orthogonal. The bottom panels show the red fractions as functions
of $M_{r}$, further demonstrating that the \redmapper{} clusters are
systematically redder than the \yang{} clusters.

\begin{figure*}
    \centering
    \includegraphics[width=0.98\textwidth]{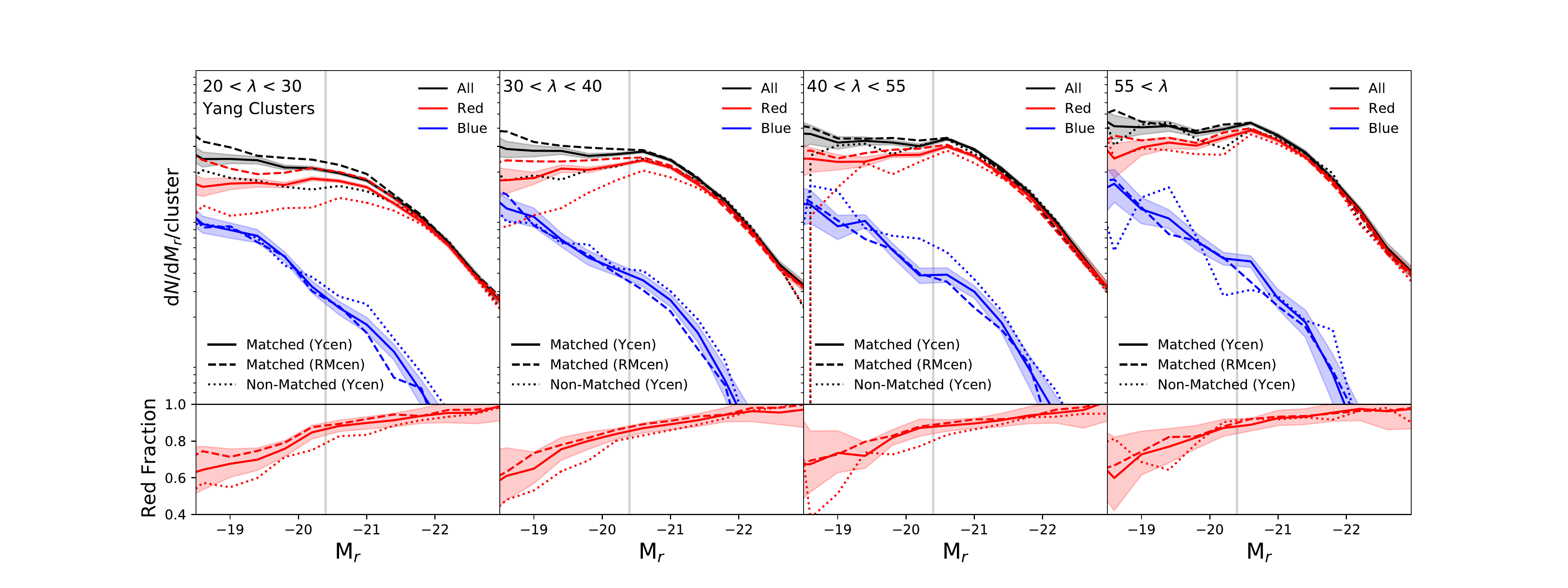}
    \caption{Comparison of the CLF shapes between the \yang{} clusters that
    have matched counterparts in \redmapper{} using luminosity-weighted
    centers~(solid curves with uncertainty bands), the same matched
    clusters using \redmapper{} centrals as centers~(dashed curves), and the
    \yang{} clusters with no matched counterparts in \redmapper{}~(dotted
    curves) in four different richness bins. In each upper panel, the
    overall CLF of each subsample~(black) is decomposed into contributions
    from the red~(red) and blue~(blue) satellite galaxies.
    The lower panels show the red fractions as functions of r-band absolute
    magnitude.} \label{fig:CLF_Yang}
\end{figure*}

Naively, one might expect the shape discrepancy between the \redmapper{}
and \yang{} CLFs shown in Figure~\ref{fig:CLF_RM_Yang} to be caused mainly
by the mismatched clusters (i.e., clusters detected in one catalogue but
not in the other).  However, from Figure~\ref{fig:overlap} we know that the
mismatched fraction rapidly decreases as a function of $\lambda$, but that
the shape discrepancies in the overall and red CLFs shown in
Figure~\ref{fig:CLF_RM_Yang} are consistent across the four richness bins.
Therefore, while the mismatched clusters are likely the culprit for the
discrepancy in the blue CLFs (the $\lambda$-dependent boost at
$M_r{<}{-}21$), the discrepancy in the red CLFs~(hence overall CLFs) is
mainly due to the different centring algorithms between the two cluster
finders. To test this hypothesis, we disentangle the effect of centring
from that of mismatching by splitting the \yang{} clusters at each
$\lambda$ bin into two subsamples, one with matched counterparts in the
\redmapper{} sample based on our fiducial cross-matching
criteria~(hereafter referred to as ``matched''), and the other one
without~(``non-matched''). In Figure~\ref{fig:CLF_Yang}, we show the
stacked CLFs of the matched clusters using both the original \yang{}
luminosity-weighted centres~(solid curves) and the highest-$\pcen$ galaxies
of their matched counterparts in \redmapper{}~(dashed curves), as well as
the stacked CLF of the non-matched clusters using their original \yang{}
centres~(dotted curves). Similar to Figure~\ref{fig:CLF_RM_Yang}, we
decompose each CLF into the contributions from red~(red curves) and
blue~(blue curves) galaxies, and show the red fractions in the bottom
panels.

\begin{figure*}
    \centering
    \includegraphics[width=0.98\textwidth]{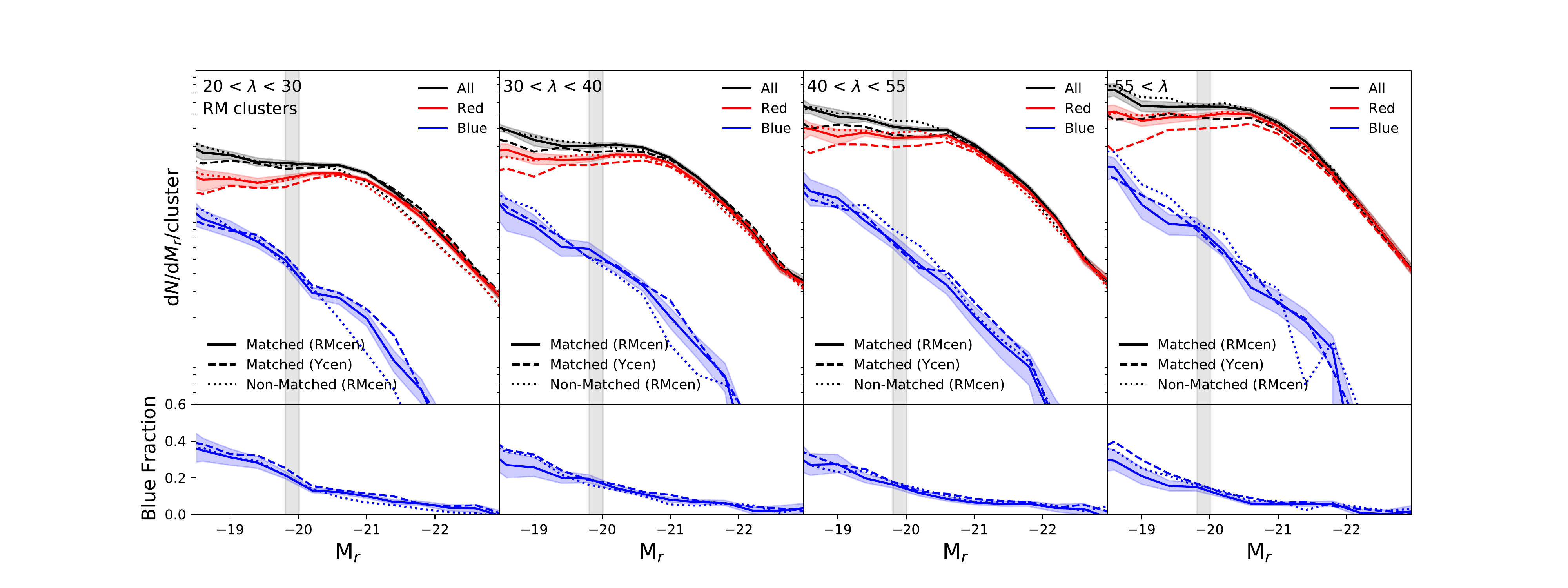}
    \caption{Similar to Figure~\ref{fig:CLF_Yang}, but for \redmapper{}
    clusters. The lower panels show the blue fractions as functions of r-band
    absolute magnitude. } \label{fig:CLF_RM}
\end{figure*}

By comparing the CLFs of the matched clusters measured with different
centres, we find that the shape discrepancy induced by switching centres
closely mimics that between the \yang{} and \redmapper{} CLFs seen in
Figure~\ref{fig:CLF_RM_Yang}. The flat (decreasing) population of all (red)
faint satellites observed around \yang{} centres~(solid black/red curves)
becomes increasing (flat) at the faint end when switching to the
\redmapper{} centrals~(dashed curves).  Therefore, we conclude that these
shape differences between the overall~(as well as red) CLFs measured for
the two cluster catalogues are primarily caused by the different centring
algorithms. The \redmapper{} algorithm tends to identify the location of
the most densely populated pocket of red galaxies as the centre of a
cluster, while the luminosity-weighted centre found by \yang{} likely sits
more closely to the geometric centre of a larger cluster region but may
miss the true centre~(defined by the minimum of the potential) by a large
margin in morphologically-disturbed systems. As a result, neither centring
algorithm works perfectly in all scenarios, and it is thus more plausible
that the true CLF of a correctly-centred cluster sample~(e.g., by X-ray
observations) is closer to the average of the two versions.  We plan to
test this hypothesis by measuring the CLF for clusters with well measured
X-ray centres in the future.

For the \yang{} clusters with no matched counterparts in
\redmapper{}~(dotted blue curves), their blue galaxy CLFs are similar to
those of the matched clusters~(solid blue
curves) in shape, but with a slightly larger amplitude at the bright end.
However, for non-matched clusters, the red galaxy
CLFs~(dotted red curves) are significantly lower than the red galaxy CLFs
of the matched clusters~(solid red curves), by an amount ranging from
20\% in the highest richness bin to a factor of two in the lowest richness
bin. Consequently, the red fraction of the non-matched clusters is
systematically lower than that of the matched clusters in all richness
bins~(bottom panels).  Since those non-matched clusters are significantly
less dominated by the red galaxies than the matched clusters, it is
unsurprising that the \redmapper{} algorithm does not detect them as
clusters with $\lambda{>}20$.  Meanwhile, the \yang{} cluster finder
detects progressively more systems with bluer satellite population than
\redmapper{}, especially in low-richness systems.

We perform a similar test using the \redmapper{} sample by splitting the
\redmapper{} clusters into matched vs. non-matched subsamples based on the
existence of matched counterparts in the \yang{} sample. The result is
shown in Figure~\ref{fig:CLF_RM}. Likewise, the shape discrepancy persists
in the matched clusters~(solid black curves), but promptly disappear after
we adopt the same centres for these clusters~(dashed black curves).  The
increasing (relatively flat) population of all (red) faint galaxies when
centered on the \redmapper{} central transforms into the relatively flat
(slightly decreasing) population of all (red) low-luminosity galaxies when
centred on the \yang{} luminosity weighted center.  Interestingly, the red
galaxy CLFs of the matched vs. non-matched clusters are consistent across
all richness bins, as the \redmapper{} algorithm regards them as
indistinguishable by counting the red galaxies.  However, the blue CLFs of
the non-matched clusters~(dotted blue curves) are much lower than those of
the matched clusters~(solid blue curves) in the two low-richness bins,
while the two sets of blue CLFs are consistent in the two high-richness
bins, mirroring the trend seen for the \yang{} sample in
Figure~\ref{fig:CLF_Yang}.  As a result, the bottom panels show that the
blue fraction of the non-matched clusters~(dotted blue curves) is generally
lower than that of the matched clusters~(solid blue curves).

To summarise the results from Figures~\ref{fig:CLF_Yang} and
~\ref{fig:CLF_RM}, the shape of the overall CLFs are primarily sensitive to
the centroiding of clusters. In particular, the faint-end slope could
exhibit an increasing or a decreasing trend, depending on whether the
cluster centres are located within a dense clump of red galaxies or a
luminosity-weighted centre averaged over a large spread of both red and
blue galaxies.  In addition, some of the discrepancies in the CLF shapes
are caused by the orthogonal colour preferences between the two cluster
finders, resulting in the non-detection of the extremely red-dominated
clusters by \yang{} and the relatively blue clusters by \redmapper{},
respectively.

\subsection{CLF Parameters and Comparison to Previous Results}
\label{subsec:params}

Since the introduction of
the Schechter Function~\citep{sch76}, the observed CLF has been
characterised by its amplitude ($\phi^{*})$, characteristic luminosity
($M^{*}$), and faint-end slope ($\alpha$), all information which provides
insight into the physical processes that characterise galaxy formation and
evolution. To measure the three parameters associated with our CLF for each
of the samples shown in Figures~\ref{fig:CLF_RM_Yang}, ~\ref{fig:CLF_Yang},
and ~\ref{fig:CLF_RM}, we fit our measured CLF to a function of the form
\begin{equation}
    \phi(M_r)=\frac{\ln(10)}{2.5}\phi^{*}10^{0.4(M^{*}-M_{r})(\alpha
    +1)}\exp[{-10^{0.4(M^{*}-M_{r})}}],
\end{equation}
using a standard Bayesian inference method using the Markov Chain Monte Carlo
(MCMC) ensemble sampler \textsc{PyMC}. The results are given in
Table~\ref{tab:CLFparam} along with the \redmapper{} results when the
central is not included. In addition, the CLF parameters for the (non-)
matched subsamples of the \yang{} and \redmapper{} clusters are
provided in the Appendix in Tables~\ref{tab:CLFparam_Yang} and
~\ref{tab:CLFparam_RM}, respectively.

\begin{table}
\centering
\caption{Best-fitting parameters of the CLFs in Figure~\ref{fig:CLF_RM_Yang} .}
\begin{tabular}{cccc}
\hline
bin & M$^{*}$ & $\alpha$ & $\phi^{*}$\\
\hline
\multicolumn{4}{|c|}{All galaxies of \yang{} clusters} \\
\hline
 1 & -22.02 $\pm$ 0.04 & -0.99 $\pm$ 0.02 & 2.44 $\pm$ 0.09 \\
 2 & -21.65 $\pm$ 0.04 & -0.83 $\pm$ 0.03 & 4.60 $\pm$ 0.16 \\
 3 & -21.64 $\pm$ 0.05 & -0.88 $\pm$ 0.03 & 5.40 $\pm$ 0.24 \\
 4 & -21.65 $\pm$ 0.06 & -0.89 $\pm$ 0.04 & 6.66 $\pm$ 0.35 \\
\hline
\multicolumn{4}{|c|}{Red galaxies of \yang{} clusters} \\
\hline
1 & -21.93 $\pm$ 0.03 & -0.83 $\pm$ 0.02 & 2.48 $\pm$ 0.07 \\
2 & -21.61 $\pm$ 0.04 & -0.70 $\pm$ 0.03 & 4.41 $\pm$ 0.14 \\
3 & -21.61 $\pm$ 0.05 & -0.77 $\pm$ 0.03 & 5.16 $\pm$ 0.22 \\
4 & -21.59 $\pm$ 0.06 & -0.78 $\pm$ 0.04 & 6.65 $\pm$ 0.34 \\
\hline
\multicolumn{4}{|c|}{Blue galaxies of \yang{} clusters} \\
\hline
1 & -21.63 $\pm$ 0.14 & -1.47 $\pm$ 0.05 & 0.29 $\pm$ 0.05 \\
2 & -21.29 $\pm$ 0.17 & -1.30 $\pm$ 0.09 & 0.55 $\pm$ 0.12 \\
3 & -21.34 $\pm$ 0.20 & -1.30 $\pm$ 0.11 & 0.58 $\pm$ 0.15 \\
4 & -21.37 $\pm$ 0.27 & -1.43 $\pm$ 0.13 & 0.52 $\pm$ 0.18 \\
\hline
\multicolumn{4}{|c|}{All galaxies of \redmapper{} clusters} \\
\hline
1 & -22.13 $\pm$ 0.04 & -1.15 $\pm$ 0.02 & 2.14 $\pm$ 0.11 \\
2 & -21.96 $\pm$ 0.05 & -1.10 $\pm$ 0.02 & 3.35 $\pm$ 0.16 \\
3 & -21.84 $\pm$ 0.05 & -1.11 $\pm$ 0.03 & 4.63 $\pm$ 0.26 \\
4 & -21.78 $\pm$ 0.06 & -1.06 $\pm$ 0.03 & 6.83 $\pm$ 0.44 \\
\hline
\multicolumn{4}{|c|}{Red galaxies of \redmapper{} clusters} \\
\hline
1 & -22.03 $\pm$ 0.04 & -1.01 $\pm$ 0.02 & 2.34 $\pm$ 0.09 \\
2 & -21.82 $\pm$ 0.04 & -0.95 $\pm$ 0.02 & 3.72 $\pm$ 0.16 \\
3 & -21.75 $\pm$ 0.05 & -0.99 $\pm$ 0.03 & 4.89 $\pm$ 0.24 \\
4 & -21.71 $\pm$ 0.05 & -0.95 $\pm$ 0.03 & 7.05 $\pm$ 0.40 \\
\hline
\multicolumn{4}{|c|}{Blue galaxies of \redmapper{} clusters} \\
\hline
1 & -21.20 $\pm$ 0.17 & -1.57 $\pm$ 0.07 & 0.32 $\pm$ 0.07 \\
2 & -21.46 $\pm$ 0.24 & -1.57 $\pm$ 0.09 & 0.32 $\pm$ 0.10 \\
3 & -21.43 $\pm$ 0.29 & -1.61 $\pm$ 0.10 & 0.38 $\pm$ 0.16 \\
4 & -21.55 $\pm$ 0.30 & -1.67 $\pm$ 0.10 & 0.36 $\pm$ 0.15 \\
\hline
\multicolumn{4}{|c|}{All galaxies of \redmapper{} clusters - no BCG} \\
\hline
1 & -21.50 $\pm$ 0.03 & -0.94 $\pm$ 0.02 & 3.64 $\pm$ 0.14 \\
2 & -21.59 $\pm$ 0.04 & -0.98 $\pm$ 0.02 & 4.54 $\pm$ 0.19 \\
3 & -21.58 $\pm$ 0.04 & -1.00 $\pm$ 0.03 & 5.83 $\pm$ 0.28 \\
4 & -21.59 $\pm$ 0.05 & -0.99 $\pm$ 0.03 & 8.01 $\pm$ 0.48 \\
\hline
\multicolumn{4}{|c|}{Red galaxies of \redmapper{} clusters - no BCG} \\
\hline
1 & -21.38 $\pm$ 0.03 & -0.75 $\pm$ 0.02 & 3.83 $\pm$ 0.11 \\
2 & -21.46 $\pm$ 0.04 & -0.80 $\pm$ 0.02 & 4.90 $\pm$ 0.17 \\
3 & -21.50 $\pm$ 0.04 & -0.89 $\pm$ 0.03 & 6.00 $\pm$ 0.25 \\
4 & -21.51 $\pm$ 0.05 & -0.87 $\pm$ 0.03 & 8.32 $\pm$ 0.44 \\
\hline
\multicolumn{4}{|c|}{Blue galaxies of \redmapper{} clusters - no BCG} \\
\hline
1 & -21.19 $\pm$ 0.17 & -1.57 $\pm$ 0.07 & 0.32 $\pm$ 0.08 \\
2 & -21.46 $\pm$ 0.24 & -1.57 $\pm$ 0.09 & 0.32 $\pm$ 0.10 \\
3 & -21.43 $\pm$ 0.29 & -1.61 $\pm$ 0.10 & 0.38 $\pm$ 0.16 \\
4 & -21.55 $\pm$ 0.29 & -1.68 $\pm$ 0.10 & 0.36 $\pm$ 0.15 \\
\end{tabular}

\label{tab:CLFparam}
\end{table}

The primary takeaway from measuring the parameters, given in
Table~\ref{tab:CLFparam}, is that only $\phi^{*}$ depends on $\lambda$,
in agreement with the results from \citet{han09}. However, the comparison
between the parameters measured for the \redmapper{} and \yang{} clusters is
more insightful. As given in Table~\ref{tab:CLFparam}, the underlying
parameters within the CLF slightly differ, as suggested by
Figure~\ref{fig:CLF_RM_Yang}. We find that the $M^{*}$ for \redmapper{} and
\yang{} clusters are largely consistent, which reflects the location
of the bright-end turn-over in Figure~\ref{fig:CLF_RM_Yang}. Moreover, as
shown in Figure~\ref{fig:CLF_RM_Yang}, we see that the faint-end
slope, $\alpha$ for the \redmapper{} catalogue is slightly steeper, for both
red, blue, and overall galaxies.  Lastly, as previously mentioned, we
unsurprisingly find that the amplitude $\phi^{*}$ is generally larger for the
\redmapper{} sample, in agreement with the results from
Figure~\ref{fig:Numgal}.

Since the galaxy cluster CLF is well studied, there are many results we can
compare our stacked measurements to. However, these parameters are
mildly covariant, so the inclusion of the central galaxy, which impacts the
measurement of M$^{*}$ also impacts $\alpha$ and $\phi$.  Based on the
inclusion of the central, we do not compare our measured M$^{*}$ to previous
results.  However, we also present the \redmapper{} results without the
central galaxy to make a comparison to the $\alpha$ measurements from
\citet{han09} and \citet{zha19b}.

One of the best comparison samples is from \citet{zha19b}, who measured the
CLFs of clusters identified by \redmapper{} in the Dark Energy Survey~(DES)
using the DES Science Verification data~\citep{ryk16}. Although this analysis
covers both a larger range in halo mass and higher redshift, we find
strong agreement between our $\alpha$ value for red galaxies when the BCG is
excluded with their measurement, $\alpha {\simeq}{-}0.7-0.8$, for DES-
\redmapper{} selected clusters in our redshift range. Of note, \citet{zha19b}
also detect a similar trend in $\alpha$, such that the value becomes slightly
steeper with richness.

We also compare our results to \citet{han09}, which cover a similar redshift
and halo mass range, but instead use the MaxBCG catalogue \citep{koe07}.
We note that \citet{han09} measure the CLF within R$_{200}$ as opposed to
a 1$\hmpc$ aperture; however, as shown in their Figure 3, their
average aperture size is comparable to ours. Another difference is that
\citet{han09} fix $\alpha$ for the blue galaxies, while we allow it to be a
free parameter. Given our use of a similar aperture, it is unsurprising
that we measure similar values $\phi^{*}$ for both the red and blue samples
and the overall sample. Additionally, the $\alpha$ value for red galaxies
($\alpha{\simeq}{-}0.7$) and overall galaxy sample ($\alpha{\simeq}{-}0.9$)
show general agreement with our results excluding the BCG, though our
values are $\sim$0.1 steeper. For both of our samples, the blue galaxy
$\alpha$ is much steeper than what \citet{han09} assume, $\alpha_{blue}{=}{-
}1.0$.  Some of these discrepancies may result from the difference in
cluster finding algorithm and how red and blue galaxies are defined.
\citet{han09} uses a fixed bimodal definition based on $g{-}r$ colours, while
we use the \redmapper{} defined red sequence which is redshift dependent.

Although measured for lower redshift galaxy clusters, we compare our
results to those presented in \citet{lan16} ($0.01<z<0.05$) and
\citet{men22} ($0.01<z<0.08$), which use clusters identified in the
\citet{yan07} spectroscopic group catalogue. The primary differences
between our measurement approaches are the different apertures within which
we measure the CLF (1$\hmpc$ vs R$_{200}$), how we identify red and
blue galaxies, and our inclusion of the central galaxy.
However, given that we both use the halo-based cluster
finder, this is a relevant comparison. We note that we only compare our
results to the bright portion of the measurements presented in \citet{lan16}
and \citet{men22}, because we do not observe the magnitude range fainter
than $M_{r}{=}{-}18$) due to the higher redshift nature of our cluster
samples.

For the total sample of galaxies, our measurements
for $\alpha$ are in agreement with the results from both \citet{lan16}
and \citet{men22}. For the red galaxies, our values of
$\alpha$ are in agreement at the lowest-$\lambda$ end,
while at the high-richness end,
our $\alpha$ is shallower by 0.15-2 than both \citet{lan16} and \citet{men22}.
For blue galaxies, we find that our measurements for
$\alpha$ are in agreement with the values presented in \citet{lan16} and
\citet{men22} for the entire $\lambda$ range. We note that the difference
between our
measurements and those from \citet{lan16} and \citet{men22} is primarily
due to the different performances of the photometric and spectroscopic
halo-based cluster finders as well as our inclusion of the central galaxy.

In addition to the CLF parameters, we also compare our measured red
fractions to those from \citet{han09} and \citet{men22}.
\citet{han09} measure a red fraction within the entire cluster of
$\sim$80\% for galaxy clusters, in strong agreement with the results we
find for both the \redmapper{} and \yang{} samples. In a more direct
comparison, \citet{men22} find that the red fraction increases slightly
with $\lambda$ and that it also increases from 60\% to 100\% with $M_{r}$
over the range $-18>M_{r}>-23$, which is in excellent agreement with
our measurement shown in Figure~\ref{fig:CLF_RM_Yang}.

In \citet{lan16}, one of their primary conclusions is that the CLF can be
characterised by a faint-end slope that steepens at
$M_{r}=-18.0$. This same faint-end slope was similarly detected by
\citet{tin21} and \citet{men22} for CLF samples measured at low redshifts
($z<0.08$). Although the higher average redshift of our sample prevents
us from directly making this measurement, we address this comparison by
extrapolating our faint-end behaviors to fainter magnitudes. In particular,
\citet{lan16} note that the increase in the faint-end slope is driven
predominately by the red galaxies, where $\alpha = -1.8$. Moreover, both
\citet{lan16} and \citet{men22} find that the fraction of red galaxies
greatly increases in the magnitude range $-12<M_{r}<-16$, and thus
results in that faint end slope. However, it is worth noting that this
trend was identified using groups and clusters identified in the
\citet{yan07} spectroscopic group catalogue, using the luminosity-weighted
centres. Interestingly, Figure~\ref{fig:CLF_RM_Yang} suggests we
may see that the faint-end population starts to modestly rise at
$M_r{=}{-}19$ for some richness bins.
However, even though we do not analyse the faintest $M_r$ range of
interest, we find that the CLF shape is sensitive to the centring
algorithm used and we expect this conclusion to hold at fainter magnitudes.
Therefore, while \citet{lan16} and \citet{men22} suggest that this steeper
faint-end slope is a result of the different formation processes behind red
and blue galaxies in the cluster environment, our results suggest that this
effect may be impacted by the choice of cluster finder, particularly the
way the centres of the galaxy clusters are identified.

To summarise the results from our stacked CLF comparison, we confirm the
expectation that the \redmapper{} cluster finder generally detects clusters
with a higher fraction of red galaxies and vice versa for the \yang{}
finder. However, the shape discrepancy between the CLFs of the two cluster
samples is, somewhat unexpectedly, mainly due to the difference in cluster
centring between the two finders. For a cluster cosmology analysis,
however, this is very encouraging news because cluster miscentring can be
robustly calibrated using X-ray data~\citep{zha19} and accurately modelled
out in the cluster weak lensing analysis when measuring halo
mass~\citep[e.g.,][]{zu21, wan22}, the key quantity for cluster cosmology.

\section{Weak Lensing Halo Mass of Clusters}
\label{sec:WL}

\subsection{Weak Lensing of Matched vs. Non-Matched Clusters}
\label{subsec:WLmotivation}

After examining the discrepancy in the galaxy content between the two
cluster samples, we now explore whether a significant systematic
uncertainty exists as a result of any cluster detection bias associated
with the galaxy content preferred by each finder. As mentioned in
Section~\ref{sec:intro}, it is plausible that the detection bias associated
with a particular cluster finder would drive the otherwise random scatter
in the mass-richness relation to be systematic, sometimes with strong
non-Gaussian tails. For example, if a cluster finder preferentially assigns
higher richnesses to haloes with a higher red fraction due to an older
formation time, while misdetecting young haloes with higher blue fraction
but the same halo mass. The scatter in the mass-richness relation would
have a systematic contribution, so that the halo mass distribution at fixed
richness would include a long tail into the low-mass regime. By the same
token, another cluster finder could assign higher richnesses to systems
with an excess number of blue galaxies projected within the 2D aperture due
to the larger photo-z errors of blue galaxies, thereby producing a similar
long tail at the low-mass end. Unfortunately, it is observationally
challenging to ascertain the existence of such a systematic component in
the scatter, as the underlying halo mass distribution of the clusters
is unknown.
\begin{table}
\centering
\caption{Parameter descriptions and priors for WL halo mass inference.}
\begin{tabular}{ccc}
\hline
Parameter & Description & Prior\\
\hline
$\log \mhwl$ & Log-halo mass & $\mathcal{U}$(13.0,16.0) \\
c & Halo concentration & $\mathcal{N}$(5,1.5$^2$) \\
$f_{\rm off,RM}$ & \redmapper{} offset fraction& $\mathcal{N}$(0.30,0.04$^2$)\\
$\sigma_{\rm off,RM}$ & \redmapper{} characteristic offset & $\mathcal{N}$(0.18,0.02$^2$) \\
    $f_{\rm off,Y}$ & \yang{} offset fraction & $\mathcal{U}$(0.0, 1.0)\\
$\sigma_{\rm off,Y}$ & \yang{} characteristic offset & $\mathcal{U}$(0.0,1.0) \\
\hline
\end{tabular}
\small
\\
$\mathcal{N}(a,b)$ refers to a Normal distribution with mean and variance of a and b and $\mathcal{U}(a,b)$ refers to a uniform distribution with upper and lower limits a and b.
\label{tab:DSbayes}
\end{table}
\begin{figure*}
    \centering
    \includegraphics[width=0.98\textwidth]{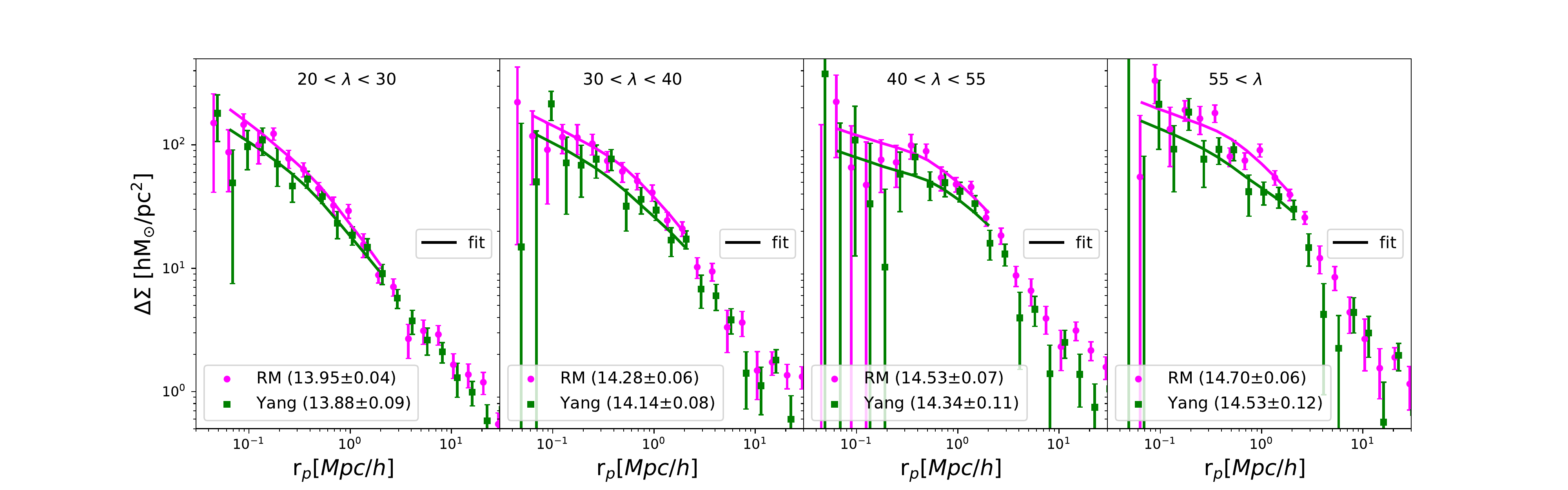}
    \caption{The $\Delta\Sigma$ profiles of the \yang{}~(green squares with
    errorbars) and \redmapper{}~(magenta circles with errorbars) clusters
    in four different richness bins. Solid green and magenta lines are the
    best-fitting model curves for the \yang{} and \redmapper{} clusters,
    respectively.} \label{fig:DS_RM_Yang}
\end{figure*}

However, by taking advantage of the two cluster samples within the same
volume detected by complementary approaches, we can circumvent the issue by
exploring whether a significant discrepancy exists in the average halo
masses estimated from weak lensing ($\mhwl$), in particular between the
matched and the non-matched clusters at fixed $\lambda$ of each cluster
catalogue. The rationale is as follows. Assuming cluster catalogues
$\mathcal{A}$ and $\mathcal{B}$ have detection biases that are roughly
orthogonal, at fixed $\lambda$ of catalogue $\mathcal{A}$, clusters that have
matched counterparts in $\mathcal{B}$ are likely the systems that are least
affected by the detection bias of $\mathcal{A}$, while the $\mathcal{A}$
clusters that do not have any matches in $\mathcal{B}$ are potentially the
ones that are intrinsically low-mass systems but promoted above the
richness/detection threshold of $\mathcal{A}$. Therefore, if we observe a
significant discrepancy in the weak lensing halo mass of the matched vs.
non-matched clusters from $\mathcal{A}$, this would suggest the existence
of a strong systematic component in the scatter, which would then lead to
severe bias in the cluster cosmology analysis.

We follow a similar approach to estimate $\mhwl$ as what is described in
depth in \citet{zu21} and use the same shear
catalogue\footnote{https://gax.sjtu.edu.cn/data/DESI.html} as in
\citet{wan22}. Therefore, here we describe the data used for our
measurements and how we fit the $\Delta\Sigma$, excess surface density
profile. A more detailed description of our cluster weak lensing formalism
is provided in the Appendix. The weak gravitational lensing signal is
measured using the weak lensing shear catalogue derived from the Dark
Energy Camera Legacy Survey (DECaLS) DR8~\citep{dey19}, which is part
of the DESI Legacy Imaging Survey. For this analysis, our source galaxy
photometry and photometric redshifts come from \citet{zou19}. These
redshifts are supplemented with the photometric redshifts estimates from
\citep{zho21}. For each richness bin, the shear catalogues are constructed
from the DECaLS DR8 data using the \textsc{Fourier\_Quad}
method~\citep{Zhang2015,Zhang2019,Zhang2022}.

\subsection{Bayesian Modelling of Cluster Weak Lensing on Small Scales}
\label{subsec:WLBayesian}

In this analysis, we use a Bayesian framework to infer the weak lensing
halo masses directly from the $\Delta\Sigma$ profiles, following the same
method as outlined in \citet{zu21}. Since our goal is to measure the halo
mass $\mhwl$, we only use the $\Delta\Sigma$ profiles measured at radii
below the transition scale between the so-called one-halo and two-halo
terms, which occurs at $\sim$2$\hmpc$~\citep{zu14}.

For each of the four $\lambda$ bins, we have four parameters that we are
fitting for the $\Delta\Sigma$ profile, including the halo mass $\mhwl$,
halo concentration $c$, fraction of miscentred clusters with offset from
the true halo centre $f_{\rm off}$, and the characteristic offset of
miscentred clusters $\sigma_{\rm off}$. We note that the miscentring
effects are highly degenerate with halo concentration if $\sigma_{\rm off}$
is allowed to freely vary. We mitigate this degeneracy for the \redmapper{}
clusters using the results from \citet{zha19} which use \textit{Chandra}
X-ray observations to constrain the offset and average miscentring
fractions as $\langle\sigma_{\rm off}\rangle$=0.18$\pm$0.02$\hmpc$ and
$\langle f_{\rm off}\rangle$=0.30$\pm$0.04. As of the time of this work, no
such analysis has been done for the \yang{} clusters, so we adopt uniform,
uninformative priors on the miscentring parameters for the $\Delta\Sigma$
of \yang{} clusters.

We determine our priors following the approach in \citet{zu21} and describe
them in Table~\ref{tab:DSbayes}. To determine the posterior distribution
for each of these four parameters, we again use the MCMC
ensemble sampler \textsc{PyMC}. We run the sampler for 500,000 steps
to ensure covariance and derive the posterior constraints after a burn-in
of 100,000 steps. The median values for
the $\mhwl$ and the 68\% confidence limits of the 1D posteriors for the halo
masses are provided in Figures~\ref{fig:DS_RM_Yang},
~\ref{fig:DS_Overlapping_Yang}, and ~\ref{fig:DS_Overlapping_RM} and the
posterior constraints on all other parameters are provided in
Tables~\ref{tab:WL_Yang} and ~\ref{tab:WL_RM} for the \yang{} and
\redmapper{} samples, respectively.

\begin{figure}
    \centering
    \includegraphics[width=0.48\textwidth]{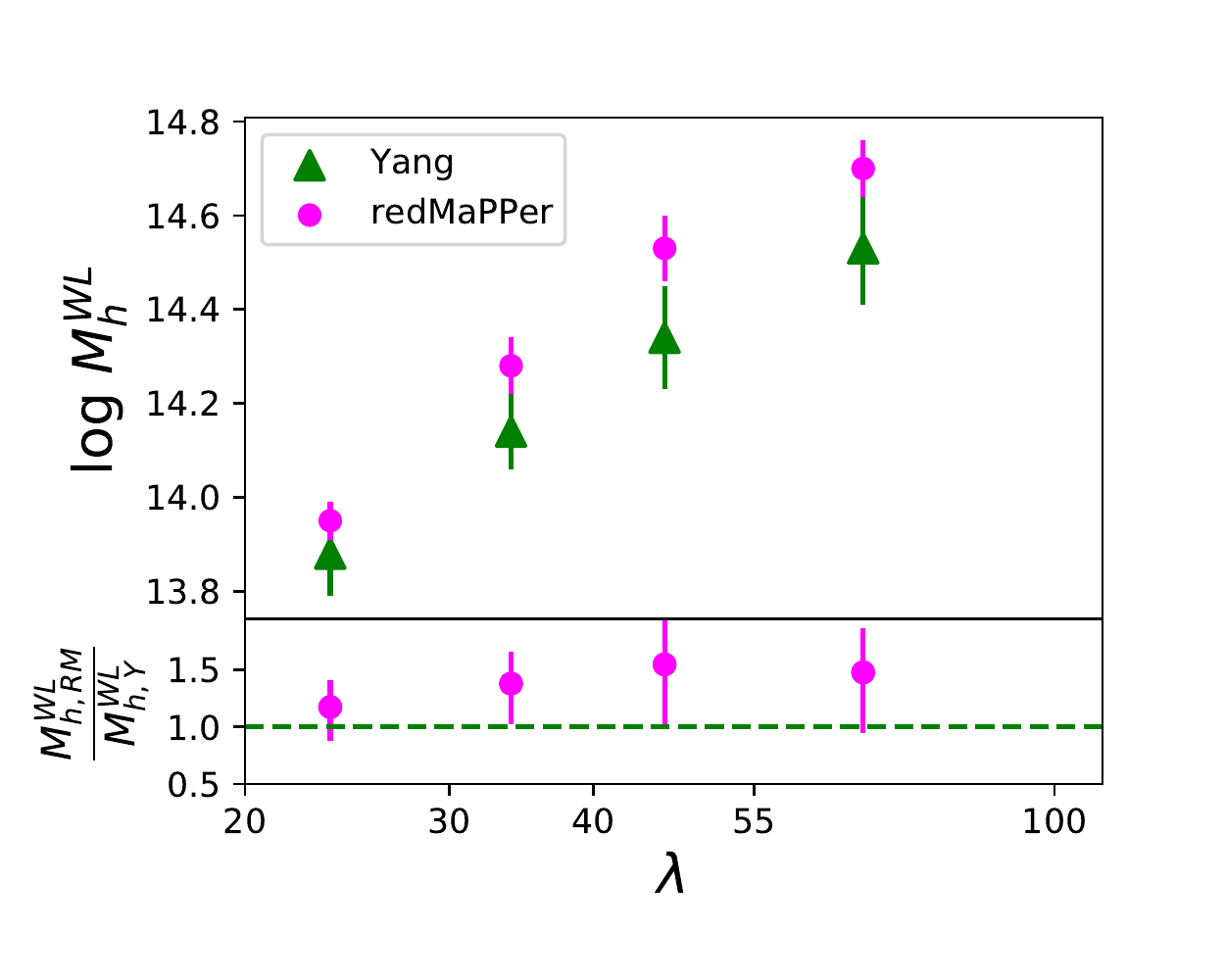}
    \caption{Similar to Figure~\ref{fig:Numgal}, but for the weak lensing
    halo masses. }
    \label{fig:MHWL}
\end{figure}

\subsection{Halo Mass Comparison between Matched vs. Non-Matched Clusters}
\label{subsec:WLMass}

We start by measuring the overall $\mhwl$-$\lambda$ relations of the
\redmapper{} and \yang{} samples. Figure~\ref{fig:DS_RM_Yang} shows the
$\Delta\Sigma$ profiles of the \redmapper{}~(magenta circles with
errorbars) and \yang{}~(green squares with errorbars) clusters, binned by
$\lambda$ in the same manner described in Section~\ref{subsec:analysis}.
In each panel, we show our fits~(solid curves) over the radial range used
from the $\Delta\Sigma$ profile to estimate the weak lensing halo mass,
which is given at the bottom of each panel and shown in
Figure~\ref{fig:MHWL}. Our $\mhwl$ estimates are consistent with the
similar measurements from \citet{wan22} and \citet{sim17} for the \yang{}
and \redmapper{} samples, respectively. Similar to Figure~\ref{fig:Numgal},
the WL halo mass of the \redmapper{} clusters is generally higher than that
of the \yang{} clusters at fixed $\lambda$, though the difference is rather
small at at $\lambda{<}30$.
Since the effect of miscentring should be largely mitigated in the $\Delta\Sigma$
modelling, the discrepancy between the two $\mhwl$-$\lambda$ relations in
Figure~\ref{fig:MHWL} should be
primarily induced by the slightly larger scatter in
our newly defined richness $\lamy$ for the \yang{} clusters compared to the
original \yang{} richness and the \redmapper{} richness.

\begin{figure*}
    \centering
    \includegraphics[width=0.98\textwidth]{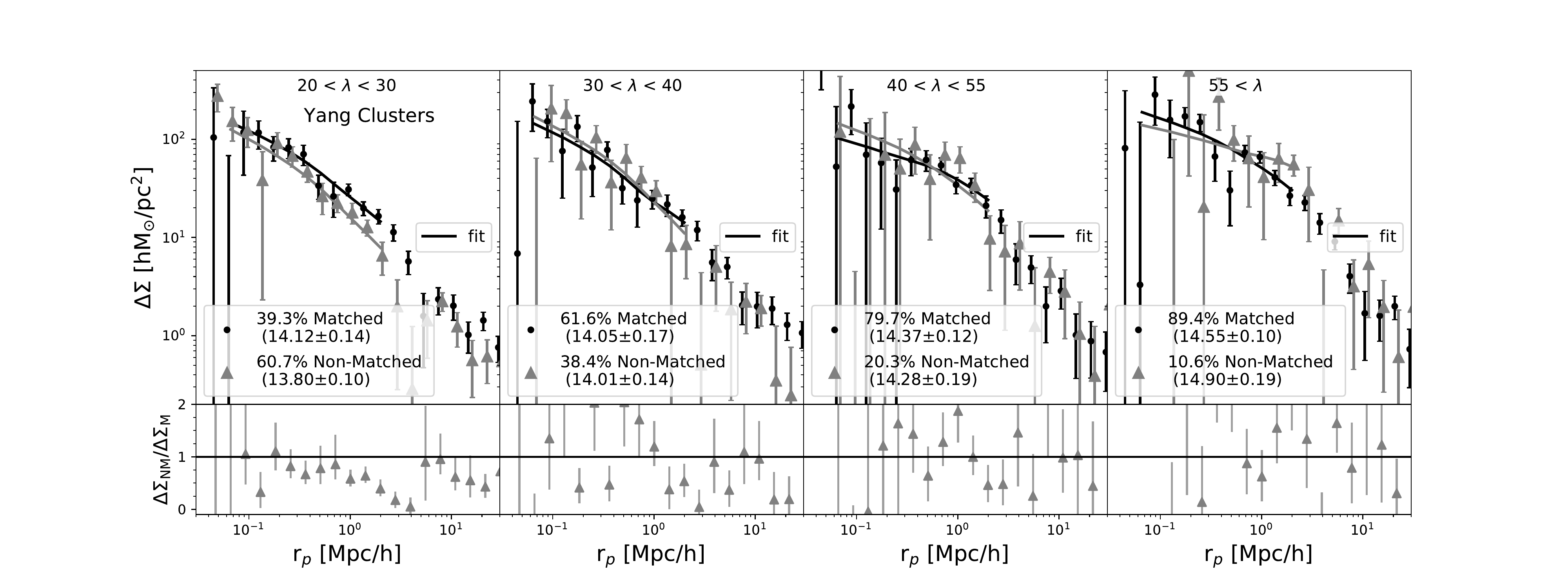}
    \caption{Comparison between the weak lensing profiles of the \yang{}
    clusters with matched counterparts in \redmapper{}~(dark circles with
    errorbars) and those without~(gray triangles with errorbars) in four
    different richness bins. Dark and gray solid curves are the
    best-fitting model curves for the matched vs. non-matched clusters,
    respectively. The lower panels show the ratio between the two
    $\Delta\Sigma$ profiles.} \label{fig:DS_Overlapping_Yang}
\end{figure*}
\begin{figure*}
    \centering
    \includegraphics[width=0.98\textwidth]{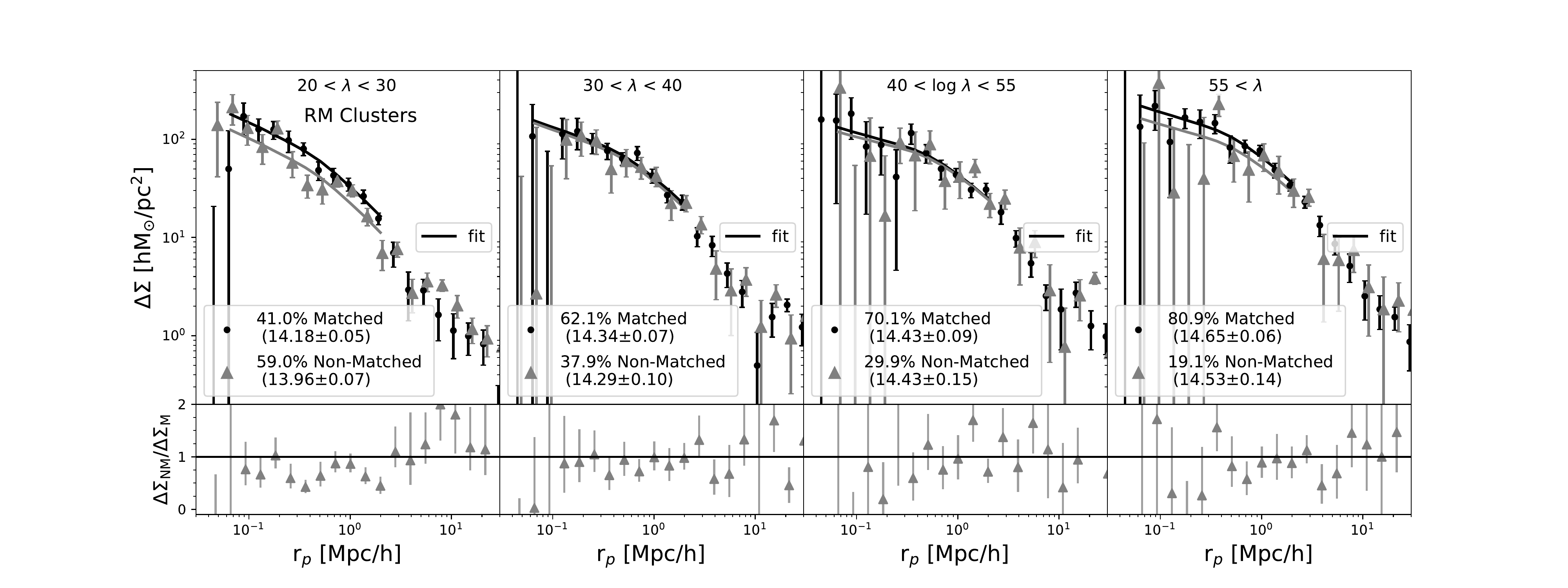}
    \caption{Similar to Figure~\ref{fig:DS_Overlapping_Yang}, but for
    \redmapper{} clusters that are split into matched vs. non-matched
    subsamples. }
    \label{fig:DS_Overlapping_RM}
\end{figure*}

We now turn to the $\Delta\Sigma$ profiles of the matched vs. non-matched
\yang{} clusters in the four bins of $\lambda$, as shown in
Figure~\ref{fig:DS_Overlapping_Yang}. In each panel, circles and triangles
are the $\Delta\Sigma$ profiles of the matched and non-matched clusters,
respectively, while dark and gray solid curves are the corresponding
best-fitting model predictions. The WL halo mass estimates and the
percentage of matched clusters within each bin are listed in the bottom of
each top panel, with each bottom panel showing the ratios between the two
$\Delta\Sigma$ profiles. The two sets of WL halo masses are
largely consistent within $1\sigma$ for \yang{} clusters above $\lambda{=}30$,
but differ by $\sim$1.4$\sigma$ in the lowest richness bin. In particular,
the average log-halo mass of the matched clusters is $14.12{\pm}0.14$, $0.32$
dex higher than that of the non-matched clusters~($13.80{\pm}0.10$). This
suggests that in the lowest-richness bin of the \yang{} catalogue, the
redder clusters are likely more massive than the systems dominated by blue
satellites. In the three higher richness bins, we do not detect any
significant discrepancy between the two subsamples. Therefore, we expect
that the \yang{} clusters should be largely free of systematic biases
induced by the detection method if limited to clusters with $\lambda>30$ 
($\mhwl{>}1.5\times10^{14}\hmsol$).

Similarly, Figure~\ref{fig:DS_Overlapping_RM} shows the result of the same
analysis using the $\Delta\Sigma$ profiles of the matched vs. non-matched
\redmapper{} clusters. Across the three largest $\lambda$ bins, the average
WL halo masses of the two subsamples are consistent with each other within
1$\sigma$, indicating that the preference of \redmapper{} to select
red-dominated clusters does not incur any significant systematic
uncertainties in the scatter for clusters with $\lambda>30$
($\mhwl{>}1.5\times10^{14}\hmsol$), but differ by $\sim$1.8$\sigma$ in the
lowest richness bin. In particular, the log-halo mass of matched clusters
is $14.18{\pm}0.05$, $0.22$ dex higher than that of the non-matched
clusters~($13.96{\pm}0.07$). This suggests that in the lowest-richness bin
of the \redmapper{} catalogue, there may exist a population of clusters
with richnesses that have been artificially inflated as a result of
projection effects in agreement with previous results
\citep[e.g.,][]{wu22, wet22}. Interestingly, in the two low-$\lambda$ bins
we find that the large-scale bias of the matched clusters is lower than
that of the non-matched ones. This bias discrepancy at fixed WL halo mass
is similar to the cluster assembly bias effect~\citep{zu21}, although it
could also be induced by a stronger projection effect in the non-matched
subsample~\citep{zu17}.

Overall, Figures~\ref{fig:DS_Overlapping_Yang} and
~\ref{fig:DS_Overlapping_RM} illustrate that when splitting the clusters
into the matched and non-matched subsamples, the two WL halo masses at
fixed $\lambda$ are always consistent with each other within $1\sigma$ for
clusters with mass above $\mhwl{\simeq}1.5\times10^{14}\hmsol$. This is encouraging
news --- despite the recognizable differences in the CLF shape between the
\yang{} and \redmapper{} cluster catalogues, our cluster weak lensing
analysis exhibits no evidence for a strong systematic uncertainty in the
scatter of the mass-richness relations in either cluster catalogue.

\section{Conclusions}
\label{sec:conclusions}

In this analysis, we focus on characterising the CLF and the WL halo masses
for galaxy clusters identified using two different methodologies, the
red-sequence~\citep{ryk14} and halo-based method~\citep{yan21}, using the
overlapping volume at $0.1{<}z{<}0.34$ where both catalogues are complete.
To mitigate the impact of different halo radius estimates, we adopt the
same aperture~($1\hmpc$) to measure the richness and the CLFs for both
cluster catalogues. In addition, we perform an abundance matching scheme
over the two catalogues, to make sure the richness distributions
of the two samples are the same above $\lambda=20$.

After thoroughly testing our stacking approach of measuring CLFs using a
mock sample of photometric galaxies, we apply the method to the
\redmapper{} and \yang{} clusters by cross-correlating clusters with the
SDSS photometric galaxy sample. We find mild differences in the total number
of member galaxies with $M_r{<}-20.4$ between the two cluster catalogues, which
is largely induced by the different random scatters in the respective
mass-richness relations.

Focusing on the shape comparison between the CLFs of the two cluster
samples at fixed $\lambda$, we observe modest differences in the behaviors
of the two samples toward the faint end; \redmapper{} clusters host more
low-luminosity red satellites with a relatively flat population and \yang{}
clusters contain a slightly declining population of fainter red galaxies.
Meanwhile, \yang{} clusters exhibit a boosted population of blue galaxies
at $M_r{<}{-}21$ compared to the \redmapper{} clusters at the low richness.
As a result, the red~(blue) fraction of the \redmapper{} clusters is
higher~(lower) than the \yang{} ones, especially at the low richness end.

To investigate the origin of the CLF shape discrepancy between the two
clusters finders, we split the clusters from each catalogue at fixed
$\lambda$ into two subsamples, one with matched counterparts in the other
catalogue and the other without. The two sets of matched subsamples retain
the same shape discrepancy exhibited by the overall sample, indicating that
the discrepancy results largely from the clusters being assigned different
centres. In particular, after we switch to the centres found by
\redmapper{} for the \yang{} clusters, we reproduce the flat/slightly
increasing faint-end slope shown by the overall \redmapper{} clusters.
Therefore, for the first time we demonstrate that the overall shape and the
faint-end slope of the CLF is sensitive to the centroiding algorithm
employed by the optical cluster finders. Therefore, a robust scheme for
mitigating miscentring is a prerequisite for any future study of the galaxy
formation physics inside clusters using CLFs, whether it be direct or
stacked measurements.

Strong colour preference in optical cluster finders does not necessarily
translate into systematic biases in the scatter of the mass-richness
relation of clusters. We investigate the latter using the same sets of
matched vs. non-matched clusters in each catalogue, by taking advantage of
the orthogonality in colour preference between the two cluster finders. In
particular, if the average halo mass of the matched clusters is
significantly different from that of the non-matched clusters at fixed
$\lambda$, we would regard it as a strong evidence for the existence of
systematic biases in the cluster catalogue. We estimate the average halo
masses using the cluster weak lensing profiles measured from the DECaLS
shear catalogue using \textsc{Fourier\_Quad} \citep{Zhang2015,Zhang2019,
Zhang2022}, by modelling the signal below $2\hmpc$ as the convolution
between a projected NFW profile and a miscentring kernel function informed
by the X-ray data~\citep{zu21}. We do not find significant discrepancies
between the weak lensing halo masses between the matched vs. non-matched
clusters at halo mass above $M_h{\simeq}1.5* 10^{14}\hmsol$ ($\lambda>30)$ for either cluster
catalogue, suggesting that any systematic trend in the scatter of the
mass-richness relation must be weak. 
However, at lower richnesses, we find discrepancies likely caused by either the
inclusion of a large population of blue galaxies or the impact of projection effects 
\citep[e.g.,][]{zu17, wu22,wet22}
Therefore, the choice of how optical
clusters are identified, whether by a red-sequence based or halo-based
approach, should not impact cluster cosmological analyses, if the clusters with 
$M_h<1.5* 10^{14}\hmsol$ are excluded.

Although we do not find strong evidence of systematic uncertainties due to
the CLF preference of cluster finders at $z{<}0.34$, it is unclear whether
detection bias would plague future cluster surveys at higher redshifts. In
particular, the Rubin Observatory Legacy Survey of Space and
Time~\citep[LSST;][]{ive19}, the China Space Station Telescope~\citep[{\it
CSST};][]{gon19}, and the Roman Space Telescope~\citep[{\it
Roman};][]{spe15} will allow us to detect clusters out to $z \approx 1.5$
\citep{ive19}. Although high-redshift clusters and proto-clusters have
been identified out to $z>1.5$~\citep[e.g.,][]{cer16,coo16,gol19b,wil20,li22} 
many of the clusters
identified at $z>1$ will look different than their low-redshift
counterparts, and not all will be characterized by a strong red sequence
\citep{bro13}. Therefore, it is important to keep pushing our finder
comparison to higher redshifts, in order to thoroughly understand the
systematics and nuances of the different methods of identifying galaxy
clusters for LSST, {\it CSST}, and {\it Roman}.

Going forward, there are many avenues of inquiry that we aim to explore
building off the analysis presented here. In the near future, we plan to
continue to investigate the CLF, including its central galaxy component, as
we investigate whether the CLF can inform BCG hierarchical growth. We plan
to split the CLF by magnitude gap, the difference in brightness between the
BCG and nth brightest cluster member. As found in \citet{gol18} the
magnitude gap is an observational proxy for hierarchical assembly so this
split may allow us to infer the properties of the progenitors of BCGs.
Additionally, we plan to perform radial splits to determine whether there
exists an observational signature of the BCG ``sphere of influence''
identified by \citet{che22} in the galaxy population. In the future, we
plan to improve the CLF and WL measurements by incorporating the
spectroscopic data from DESI~\citep{Abareshi2022} and centring information
from eROSITA~\citep{Predehl2021}, thereby allowing us to better constrain
cluster identification and properties as we enter the era of {\it CSST},
LSST, and {\it Roman}.

\section*{Data availability}

The data underlying this article will be shared on reasonable request to the corresponding author.

\section*{Acknowledgements}

We thank Emmet Golden-Marx, Yuanyuan Zhang, and Zhiwei Shao for stimulating
discussions about this work. JGM and YZ acknowledge the support by the
National Key Basic Research and Development Program of China (No.
2018YFA0404504), the National Science Foundation of China (11873038,
11621303, 11890692, 12173024), the science research grants from the China
Manned Space Project (No. CMS-CSST-2021-A01, CMS-CSST-2021-A02,
CMS-CSST-2021-B01). JZ acknowledges the support by the National Science
Foundation of China (11621303, 11890691, 12073017). YZ, XHY, and JZ
acknowledge the support from the ``111'' project of the Ministry of
Education under grant No. B20019. YZ ackonwleges the generous sponsorship
from Yangyang Development Fund, and thanks Cathy Huang for her hospitality
during the pandemic at the Zhangjiang Hi-Technology Park where he worked on
this project.

\appendix
\section{CLF Parameters of Matched vs. Non-Matched Subsamples}
\label{sec:appendix}

In Table~\ref{tab:CLFparam_Yang}, we present the CLF parameters for
the \yang{} subsamples with and without a cross-match. In
Table~\ref{tab:CLFparam_RM}, we present the CLF parameters for the
\redmapper{} subsamples with and without a cross-match.

\begin{table}
\centering
\caption{Best-fitting CLF parameters of the \yang{} cluster subsamples in
    Figure~\ref{fig:CLF_Yang}. Ycen or RMcen indicates that the \yang{} or
    \redmapper{} centres are adopted for the CLF measurement. }
\begin{tabular}{cccc}
\hline
bin & M$^{*}$ & $\alpha$ & $\phi^{*}$\\
\hline
\multicolumn{4}{|c|}{All galaxies of Matched \yang{} clusters (Ycen)} \\
\hline
1 & -22.05 $\pm$ 0.06 & -1.07 $\pm$ 0.03 & 2.42 $\pm$ 0.13 \\
2 & -21.73 $\pm$ 0.05 & -0.91 $\pm$ 0.03 & 4.28 $\pm$ 0.21 \\
3 & -21.63 $\pm$ 0.06 & -0.87 $\pm$ 0.04 & 5.49 $\pm$ 0.29 \\
4 & -21.65 $\pm$ 0.06 & -0.90 $\pm$ 0.04 & 6.68 $\pm$ 0.40 \\
\hline
\multicolumn{4}{|c|}{Red galaxies of Matched \yang{} clusters (Ycen)} \\
\hline
1 & -21.95 $\pm$ 0.05 & -0.93 $\pm$ 0.03 & 2.58 $\pm$ 0.12 \\
2 & -21.65 $\pm$ 0.05 & -0.78 $\pm$ 0.03 & 4.28 $\pm$ 0.18 \\
3 & -21.62 $\pm$ 0.06 & -0.79 $\pm$ 0.04 & 5.15 $\pm$ 0.25 \\
4 & -21.60 $\pm$ 0.06 & -0.79 $\pm$ 0.04 & 6.56 $\pm$ 0.36 \\
\hline
\multicolumn{4}{|c|}{Blue galaxies of Matched \yang{} clusters (Ycen)} \\
\hline
1 & -21.62 $\pm$ 0.23 & -1.56 $\pm$ 0.08 & 0.24 $\pm$ 0.07 \\
2 & -21.48 $\pm$ 0.25 & -1.42 $\pm$ 0.11 & 0.40 $\pm$ 0.13 \\
3 & -21.36 $\pm$ 0.23 & -1.33 $\pm$ 0.12 & 0.54 $\pm$ 0.15 \\
4 & -21.22 $\pm$ 0.30 & -1.35 $\pm$ 0.16 & 0.65 $\pm$ 0.24 \\
\hline
\multicolumn{4}{|c|}{All galaxies of Matched \yang{} clusters (RMcen)} \\
\hline
1 & -22.28 $\pm$ 0.08 & -1.21 $\pm$ 0.04 & 1.95 $\pm$ 0.18 \\
2 & -21.91 $\pm$ 0.07 & -1.07 $\pm$ 0.04 & 3.43 $\pm$ 0.26 \\
3 & -21.79 $\pm$ 0.07 & -1.01 $\pm$ 0.04 & 4.43 $\pm$ 0.32 \\
4 & -21.72 $\pm$ 0.08 & -0.97 $\pm$ 0.05 & 5.97 $\pm$ 0.53 \\
\hline
\multicolumn{4}{|c|}{Red galaxies of Matched \yang{} clusters (RMcen)} \\
\hline
1 & -22.21 $\pm$ 0.07 & -1.11 $\pm$ 0.04 & 2.05 $\pm$ 0.17 \\
2 & -21.86 $\pm$ 0.06 & -0.95 $\pm$ 0.04 & 3.50 $\pm$ 0.23 \\
3 & -21.79 $\pm$ 0.07 & -0.94 $\pm$ 0.04 & 4.16 $\pm$ 0.29 \\
4 & -21.63 $\pm$ 0.08 & -0.86 $\pm$ 0.05 & 6.11 $\pm$ 0.49 \\
\hline
\multicolumn{4}{|c|}{Blue galaxies of Matched \yang{} clusters (RMcen)} \\
\hline
1 & -21.51 $\pm$ 0.31 & -1.62 $\pm$ 0.11 & 0.23 $\pm$ 0.10 \\
2 & -21.25 $\pm$ 0.28 & -1.43 $\pm$ 0.13 & 0.44 $\pm$ 0.16 \\
3 & -21.32 $\pm$ 0.25 & -1.41 $\pm$ 0.12 & 0.46 $\pm$ 0.15 \\
4 & -21.58 $\pm$ 0.36 & -1.54 $\pm$ 0.14 & 0.35 $\pm$ 0.18 \\
\hline
\multicolumn{4}{|c|}{All galaxies of Non-Matched \yang{} clusters} \\
\hline
1 & -21.96 $\pm$ 0.05 & -0.90 $\pm$ 0.03 & 2.52 $\pm$ 0.11 \\
2 & -21.54 $\pm$ 0.06 & -0.69 $\pm$ 0.05 & 4.87 $\pm$ 0.25 \\
3 & -21.65 $\pm$ 0.12 & -0.85 $\pm$ 0.08 & 5.33 $\pm$ 0.60 \\
4 & -21.57 $\pm$ 0.22 & -0.79 $\pm$ 0.17 & 6.70 $\pm$ 1.35 \\
\hline
\multicolumn{4}{|c|}{Red galaxies of Non-Matched \yang{} clusters} \\
\hline
1 & -21.86 $\pm$ 0.04 & -0.72 $\pm$ 0.03 & 2.53 $\pm$ 0.09 \\
2 & -21.50 $\pm$ 0.06 & -0.52 $\pm$ 0.05 & 4.54 $\pm$ 0.20 \\
3 & -21.59 $\pm$ 0.12 & -0.69 $\pm$ 0.09 & 5.02 $\pm$ 0.49 \\
4 & -21.39 $\pm$ 0.22 & -0.61 $\pm$ 0.18 & 7.11 $\pm$ 1.18 \\
\hline
\multicolumn{4}{|c|}{Blue galaxies of Non-Matched \yang{} clusters} \\
\hline
1 & -21.67 $\pm$ 0.17 & -1.44 $\pm$ 0.06 & 0.31 $\pm$ 0.06 \\
2 & -20.98 $\pm$ 0.22 & -1.03 $\pm$ 0.17 & 0.87 $\pm$ 0.21 \\
3 & -21.14 $\pm$ 0.44 & -1.21 $\pm$ 0.27 & 0.93 $\pm$ 0.45 \\
4 & -20.94 $\pm$ 0.96 & -1.03 $\pm$ 0.68 & 0.75 $\pm$ 1.20 \\
\hline
\end{tabular}
\label{tab:CLFparam_Yang}
\end{table}

\begin{table}
\centering
\caption{Best-fitting CLF parameters of the \redmapper{} cluster subsamples
in Figure~\ref{fig:CLF_RM}. Ycen or RMcen indicates that the \yang{} or
    \redmapper{} centres are adopted for the CLF measurement. }
\begin{tabular}{cccc}
\hline
bin & M$^{*}$ & $\alpha$ & $\phi^{*}$\\
\hline
\multicolumn{4}{|c|}{All galaxies of Matched \redmapper{} clusters (RMcen)} \\
\hline
1 & -22.00 $\pm$ 0.05 & -1.06 $\pm$ 0.03 & 2.71 $\pm$ 0.15 \\
2 & -21.93 $\pm$ 0.05 & -1.08 $\pm$ 0.02 & 3.57 $\pm$ 0.20 \\
3 & -21.85 $\pm$ 0.06 & -1.09 $\pm$ 0.03 & 4.63 $\pm$ 0.29 \\
4 & -21.77 $\pm$ 0.06 & -1.05 $\pm$ 0.03 & 6.85 $\pm$ 0.48 \\
\hline
\multicolumn{4}{|c|}{Red galaxies of Matched \redmapper{} clusters (RMcen)} \\
\hline
1 & -21.94 $\pm$ 0.05 & -0.94 $\pm$ 0.03 & 2.77 $\pm$ 0.13 \\
2 & -21.86 $\pm$ 0.05 & -0.96 $\pm$ 0.03 & 3.66 $\pm$ 0.18 \\
3 & -21.75 $\pm$ 0.06 & -0.97 $\pm$ 0.03 & 4.99 $\pm$ 0.29 \\
4 & -21.72 $\pm$ 0.06 & -0.95 $\pm$ 0.03 & 7.00 $\pm$ 0.45 \\
\hline
\multicolumn{4}{|c|}{Blue galaxies of Matched \redmapper{} clusters (RMcen)} \\
\hline
1 & -21.57 $\pm$ 0.26 & -1.56 $\pm$ 0.10 & 0.25 $\pm$ 0.09 \\
2 & -21.23 $\pm$ 0.26 & -1.41 $\pm$ 0.13 & 0.48 $\pm$ 0.16 \\
3 & -21.56 $\pm$ 0.33 & -1.67 $\pm$ 0.11 & 0.30 $\pm$ 0.14 \\
4 & -21.44 $\pm$ 0.33 & -1.59 $\pm$ 0.12 & 0.42 $\pm$ 0.19 \\
\hline
\multicolumn{4}{|c|}{All galaxies of Matched \redmapper{} clusters (Ycen)} \\
\hline
1 & -21.93 $\pm$ 0.04 & -0.96 $\pm$ 0.02 & 3.12 $\pm$ 0.13 \\
2 & -21.91 $\pm$ 0.05 & -0.99 $\pm$ 0.03 & 3.74 $\pm$ 0.19 \\
3 & -21.74 $\pm$ 0.06 & -0.99 $\pm$ 0.03 & 5.09 $\pm$ 0.30 \\
4 & -21.58 $\pm$ 0.07 & -0.93 $\pm$ 0.04 & 7.36 $\pm$ 0.48 \\
\hline
\multicolumn{4}{|c|}{Red galaxies of Matched \redmapper{} clusters (Ycen)} \\
\hline
1 & -21.84 $\pm$ 0.04 & -0.81 $\pm$ 0.02 & 3.22 $\pm$ 0.12 \\
2 & -21.83 $\pm$ 0.05 & -0.87 $\pm$ 0.03 & 3.81 $\pm$ 0.16 \\
3 & -21.68 $\pm$ 0.05 & -0.87 $\pm$ 0.03 & 5.14 $\pm$ 0.26 \\
4 & -21.45 $\pm$ 0.07 & -0.77 $\pm$ 0.04 & 7.85 $\pm$ 0.45 \\
\hline
\multicolumn{4}{|c|}{Blue galaxies of Matched \redmapper{} clusters (Ycen)} \\
\hline
1 & -21.46 $\pm$ 0.18 & -1.42 $\pm$ 0.08 & 0.36 $\pm$ 0.08 \\
2 & -21.43 $\pm$ 0.25 & -1.44 $\pm$ 0.11 & 0.41 $\pm$ 0.13 \\
3 & -21.68 $\pm$ 0.30 & -1.56 $\pm$ 0.11 & 0.33 $\pm$ 0.13 \\
4 & -21.40 $\pm$ 0.29 & -1.56 $\pm$ 0.11 & 0.48 $\pm$ 0.19 \\
\hline
\multicolumn{4}{|c|}{All galaxies of Non-Matched \redmapper{} clusters} \\
\hline
1 & -22.15 $\pm$ 0.05 & -1.19 $\pm$ 0.02 & 1.92 $\pm$ 0.12 \\
2 & -21.96 $\pm$ 0.08 & -1.11 $\pm$ 0.03 & 3.25 $\pm$ 0.26 \\
3 & -21.75 $\pm$ 0.09 & -1.10 $\pm$ 0.04 & 4.91 $\pm$ 0.47 \\
4 & -21.75 $\pm$ 0.11 & -1.08 $\pm$ 0.05 & 6.90 $\pm$ 0.76 \\
\hline
\multicolumn{4}{|c|}{Red galaxies of Non-Matched \redmapper{} clusters} \\
\hline
1 & -22.06 $\pm$ 0.05 & -1.05 $\pm$ 0.02 & 2.13 $\pm$ 0.11 \\
2 & -21.80 $\pm$ 0.07 & -0.95 $\pm$ 0.04 & 3.67 $\pm$ 0.25 \\
3 & -21.64 $\pm$ 0.08 & -0.97 $\pm$ 0.04 & 5.19 $\pm$ 0.55 \\
4 & -21.64 $\pm$ 0.10 & -0.95 $\pm$ 0.05 & 7.40 $\pm$ 0.93 \\
\hline
\multicolumn{4}{|c|}{Blue galaxies of Non-Matched \redmapper{} clusters} \\
\hline
1 & -20.75 $\pm$ 0.24 & -1.51 $\pm$ 0.11 & 0.49 $\pm$ 0.16 \\
2 & -21.43 $\pm$ 0.50 & -1.67 $\pm$ 0.13 & 0.27 $\pm$ 0.18 \\
3 & -21.00 $\pm$ 0.43 & -1.39 $\pm$ 0.19 & 0.77 $\pm$ 0.44 \\
4 & -21.00 $\pm$ 0.64 & -1.59 $\pm$ 0.24 & 0.78 $\pm$ 0.70 \\
\hline
\end{tabular}
\label{tab:CLFparam_RM}
\end{table}

\section{Theoretical Model of $\Delta\Sigma$}

In this analysis, we use a Bayesian framework to infer $\mhwl$ directly
from the $\Delta\Sigma$ profiles, following the same mathematical formalism
as outlined in \citet{zu21}, which are computed as
\begin{equation}
    \Delta\Sigma(r_{p}) = \overline{\Sigma}({<}r_{p}) - \Sigma(r_{p}),
\end{equation}
where $\overline{\Sigma}({<}r_{p})$ and $\Sigma(r_{p})$ are the average
surface matter density within and at the projected radius r$_{p}$. Without
miscentring, $\Sigma(r_{p})$ can be computed from the 3D isotropic
halo-matter cross-correlation function $\xi_{\rm hm}(r)$,
\begin{equation}
    \Sigma(r_{p}) = \rho_{m} \int_{-\infty}^{+\infty} \xi_{\rm hm}(r_{p},r_{\pi}) dr_{\pi},
\end{equation}
where $\rho_{m}$ is the mean density of the Universe.

To model the effect of miscentring, we assume that the fraction of miscentred
BCGs is $f_{\rm off}$ with an offset of $\sigma_{\rm off}$ from the true cluster centre and that this distribution follows a shape-2 Gamma distribution $p(r_{\rm off})$ with a characteristic offset $\sigma_{\rm off}$,
\begin{equation}
    p(r_{\rm off}) = \frac{r_{\rm off}}{\sigma_{\rm off}^{2}} \exp\left(-
    \frac{r_{\rm off}}{\sigma_{\rm off}}\right).
\end{equation}
Thus, the observed surface matter density profile when accounting for miscentring can be written as
\begin{equation}
    \Sigma^{\rm obs}(r_{p}) = f_{\rm off}\Sigma^{\rm off}(r_{p}) + (1-f_{\rm off})\Sigma(r_{p}),
\end{equation}
where
\begin{equation}
\begin{split}\label{eq:6}
    \Sigma^{\rm off}(r_{p}) = & \frac{1}{2\pi}\int_{0}^{\infty} dr_{\rm off}p(r_{\rm off}) \\
    & \times \int_{0}^{2\pi}d\theta\Sigma(\sqrt{r_{p}^2 + r_{\rm off}^2 - 2r_{p}r_{\rm off}^{2}}\cos{\theta})
\end{split}\\
\end{equation}
The purpose of this weak lensing analysis is to compare the average halo
masses of different cluster samples, which are primarily inferred from
$\Delta\Sigma$ on scales below $2\hmpc$~\citep{zu14}. Therefore, we
only model $\xi_{\rm 1h}$, the ``1-halo'' term of $\xi_{\rm hm}$, for
predicting $\Delta\Sigma$ at $r_p{<}2\hmpc$~\citet{zu14}
\begin{equation}
    \xi_{\rm 1h}=\frac{\rho_{\mathrm{NFW}}(r|M_{h})}{\rho_{m}}-1,
\end{equation}
where $\rho_{\mathrm{NFW}}$ is the NFW density profile. Additionally, we
fit to the observed $\Delta\Sigma$ down to a minimum scale of
$r_p=0.1\hmpc$, below which the extra lensing effect caused by the stellar
mass of the BCG needs to be incorporated.

\subsection{WL Parameters}

In Tables~\ref{tab:WL_Yang} and \ref{tab:WL_RM}, we present the 1D posterior
distributions for the fitting to the $\Delta\Sigma$ profiles used to
estimate $\mhwl$.
\begin{table}
\centering
\caption{Best-fitting WL model parameters of the \yang{} cluster subsamples in Figures~\ref{fig:DS_RM_Yang} and ~\ref{fig:DS_Overlapping_Yang}.}
\begin{tabular}{ccccc}
\hline
bin & log $\mhwl$ & c & $\sigma_\mathrm{off}$ & f$_\mathrm{off}$\\
\hline
\multicolumn{5}{|c|}{Overall \yang{} clusters} \\
\hline
1 & 13.89 $\pm$ 0.09 & 5.06 $\pm$ 1.15 & 0.43 $\pm$ 0.30 & 0.23 $\pm$ 0.13 \\
2 & 14.14 $\pm$ 0.08 & 4.58 $\pm$ 1.30 & 0.43 $\pm$ 0.31 & 0.35 $\pm$ 0.18 \\
3 & 14.34 $\pm$ 0.11 & 4.08 $\pm$ 1.52 & 0.31 $\pm$ 0.25 & 0.52 $\pm$ 0.24 \\
4 & 14.53 $\pm$ 0.12 & 4.79 $\pm$ 1.27 & 0.50 $\pm$ 0.31 & 0.42 $\pm$ 0.17 \\
\hline
\multicolumn{5}{|c|}{Matched \yang{} clusters} \\
\hline
1 & 14.12 $\pm$ 0.14 & 5.18 $\pm$ 1.21 & 0.52 $\pm$ 0.29 & 0.31 $\pm$ 0.19 \\
2 & 14.05 $\pm$ 0.17 & 5.36 $\pm$ 1.29 & 0.52 $\pm$ 0.29 & 0.31 $\pm$ 0.19 \\
3 & 14.37 $\pm$ 0.12 & 4.50 $\pm$ 1.46 & 0.26 $\pm$ 0.25 & 0.52 $\pm$ 0.22 \\
4 & 14.55 $\pm$ 0.10 & 4.00 $\pm$ 1.03 & 0.42 $\pm$ 0.26 & 0.34 $\pm$ 0.15 \\
\hline
\multicolumn{5}{|c|}{Non-Matched \yang{} clusters} \\
\hline
1 & 13.80 $\pm$ 0.10 & 5.02 $\pm$ 1.22 & 0.45 $\pm$ 0.30 & 0.21 $\pm$ 0.14 \\
2 & 14.01 $\pm$ 0.14 & 5.31 $\pm$ 1.20 & 0.52 $\pm$ 0.31 & 0.17 $\pm$ 0.15 \\
3 & 14.28 $\pm$ 0.19 & 4.50 $\pm$ 1.36 & 0.39 $\pm$ 0.29 & 0.29 $\pm$ 0.21 \\
4 & 14.90 $\pm$ 0.19 & 4.76 $\pm$ 1.36 & 0.49 $\pm$ 0.30 & 0.59 $\pm$ 0.23 \\
\hline
\end{tabular}
\label{tab:WL_Yang}
\end{table}

\begin{table}
\centering
\caption{Best-fitting WL model parameters of the \redmapper{} cluster samples in Figures~\ref{fig:DS_RM_Yang} and ~\ref{fig:DS_Overlapping_RM}.}
\begin{tabular}{ccccc}
\hline
bin & $\log\mhwl$ & c & $\sigma_\mathrm{off}$ & f$_\mathrm{off}$\\
\hline
\multicolumn{5}{|c|}{Overall \redmapper{} clusters} \\
\hline
1 & 13.95 $\pm$ 0.04 & 7.09 $\pm$ 0.90 & 0.18 $\pm$ 0.02 & 0.27 $\pm$ 0.04 \\
2 & 14.28 $\pm$ 0.06 & 5.24 $\pm$ 0.88 & 0.18 $\pm$ 0.02 & 0.30 $\pm$ 0.04 \\
3 & 14.53 $\pm$ 0.07 & 3.65 $\pm$ 0.75 & 0.18 $\pm$ 0.02 & 0.31 $\pm$ 0.04 \\
4 & 14.70 $\pm$ 0.06 & 4.88 $\pm$ 0.82 & 0.18 $\pm$ 0.02 & 0.17 $\pm$ 0.04 \\
\hline
\multicolumn{5}{|c|}{Matched \redmapper{} clusters} \\
\hline
1 & 14.18 $\pm$ 0.05 & 5.81 $\pm$ 0.90 & 0.18 $\pm$ 0.02 & 0.30 $\pm$ 0.04 \\
2 & 14.34 $\pm$ 0.07 & 4.62 $\pm$ 0.95 & 0.18 $\pm$ 0.02 & 0.30 $\pm$ 0.04 \\
3 & 14.43 $\pm$ 0.09 & 3.84 $\pm$ 1.01 & 0.18 $\pm$ 0.02 & 0.31 $\pm$ 0.04 \\
4 & 14.65 $\pm$ 0.06 & 4.98 $\pm$ 0.84 & 0.18 $\pm$ 0.02 & 0.30 $\pm$ 0.04 \\
\hline
\multicolumn{5}{|c|}{Non-Matched \redmapper{} clusters} \\
\hline
1 & 13.96 $\pm$ 0.07 & 5.00 $\pm$ 0.94 & 0.18 $\pm$ 0.02 & 0.30 $\pm$ 0.04 \\
2 & 14.29 $\pm$ 0.10 & 4.61 $\pm$ 1.16 & 0.18 $\pm$ 0.18 & 0.31 $\pm$ 0.04 \\
3 & 14.43 $\pm$ 0.15 & 3.51 $\pm$ 1.10 & 0.18 $\pm$ 0.02 & 0.28 $\pm$ 0.04 \\
4 & 14.53 $\pm$ 0.14 & 5.40 $\pm$ 1.02 & 0.18 $\pm$ 0.02 & 0.30 $\pm$ 0.04 \\
\hline
\end{tabular}
\label{tab:WL_RM}
\end{table}

\label{lastpage}

\end{document}